\documentclass[twoside,twocolumn,9pt]{article}
\usepackage{extsizes}
\usepackage[super,sort&compress,comma]{natbib} 
\usepackage[version=3]{mhchem}
\usepackage[left=1.5cm, right=1.5cm, top=1.785cm, bottom=2.0cm]{geometry}
\usepackage{balance}
\usepackage{mathptmx}
\usepackage{sectsty}
\usepackage{graphicx} 
\usepackage{lastpage}
\usepackage[format=plain,justification=justified,singlelinecheck=false,font={stretch=1.125,small,sf},labelfont=bf,labelsep=space]{caption}
\usepackage{float}
\usepackage{fancyhdr}
\usepackage{fnpos}
\usepackage{subcaption}
\usepackage{ifthen}
\usepackage[usenames,dvipsnames]{xcolor}
\usepackage[utf8]{inputenc}
\usepackage{newunicodechar}
\newunicodechar{−}{\textminus}

\newboolean{showhighlights}
\setboolean{showhighlights}{true} 
\definecolor{zblue}{RGB}{14, 100, 190} %

\ifthenelse{\boolean{showhighlights}}%
  {}%
  {}

\usepackage[english]{babel}
\addto{\captionsenglish}{%
  
}
\usepackage{array}
\usepackage{charter}
\usepackage[T1]{fontenc}
\usepackage{setspace}
\usepackage[compact]{titlesec}
\usepackage{hyperref}

\usepackage{epstopdf}

\definecolor{cream}{RGB}{222,217,201}

\begin{document}

\pagestyle{fancy}
\thispagestyle{plain}
\fancypagestyle{plain}{
\renewcommand{\headrulewidth}{0pt}
}

\makeFNbottom
\makeatletter
\renewcommand\LARGE{\@setfontsize\LARGE{15pt}{17}}
\renewcommand\Large{\@setfontsize\Large{12pt}{14}}
\renewcommand\large{\@setfontsize\large{10pt}{12}}
\renewcommand\footnotesize{\@setfontsize\footnotesize{7pt}{10}}
\makeatother

\renewcommand{\thefootnote}{\fnsymbol{footnote}}
\renewcommand\footnoterule{\vspace*{1pt}%
\color{cream}\hrule width 3.5in height 0.4pt \color{black}\vspace*{5pt}} 
\setcounter{secnumdepth}{5}

\makeatletter 
\renewcommand\@biblabel[1]{#1}            
\renewcommand\@makefntext[1]%
{\noindent\makebox[0pt][r]{\@thefnmark\,}#1}
\makeatother 
\renewcommand{\figurename}{\small{Fig.}~}
\sectionfont{\sffamily\Large}
\subsectionfont{\normalsize}
\subsubsectionfont{\bf}
\setstretch{1.125} 
\setlength{\skip\footins}{0.8cm}
\setlength{\footnotesep}{0.25cm}
\setlength{\jot}{10pt}
\titlespacing*{\section}{0pt}{4pt}{4pt}
\titlespacing*{\subsection}{0pt}{15pt}{1pt}

\fancyfoot{}
\fancyfoot[LO,RE]{\vspace{-7.1pt}\includegraphics[height=9pt]{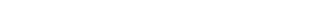}}
\fancyfoot[CO]{\vspace{-7.1pt}\hspace{11.9cm}\includegraphics{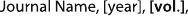}}
\fancyfoot[CE]{\vspace{-7.2pt}\hspace{-13.2cm}\includegraphics{head_foot/RF}}
\fancyfoot[RO]{\footnotesize{\sffamily{1--\pageref{LastPage} ~\textbar  \hspace{2pt}\thepage}}}
\fancyfoot[LE]{\footnotesize{\sffamily{\thepage~\textbar\hspace{4.65cm} 1--\pageref{LastPage}}}}
\fancyhead{}
\renewcommand{\headrulewidth}{0pt} 
\renewcommand{\footrulewidth}{0pt}
\setlength{\arrayrulewidth}{1pt}
\setlength{\columnsep}{6.5mm}
\setlength\bibsep{1pt}

\makeatletter 
\newlength{\figrulesep} 
\setlength{\figrulesep}{0.5\textfloatsep} 

\newcommand{\topfigrule}{\vspace*{-1pt}%
\noindent{\color{cream}\rule[-\figrulesep]{\columnwidth}{1.5pt}} }

\newcommand{\botfigrule}{\vspace*{-2pt}%
\noindent{\color{cream}\rule[\figrulesep]{\columnwidth}{1.5pt}} }

\newcommand{\dblfigrule}{\vspace*{-1pt}%
\noindent{\color{cream}\rule[-\figrulesep]{\textwidth}{1.5pt}} }

\makeatother

\twocolumn[
  \begin{@twocolumnfalse}
{\includegraphics[height=30pt]{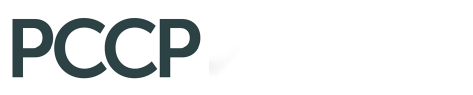}\hfill\raisebox{0pt}[0pt][0pt]{\includegraphics[height=55pt]{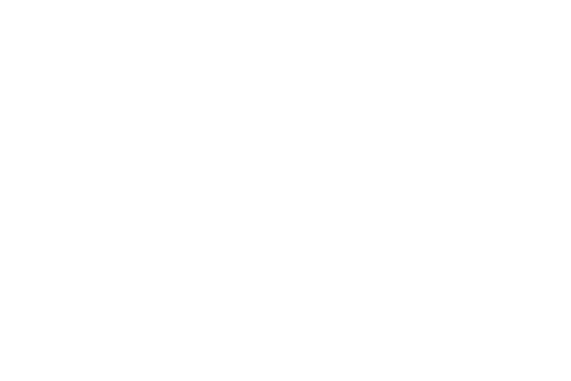}}\\[1ex]
\includegraphics[width=18.5cm]{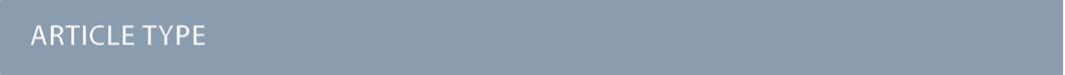}}\par
\vspace{1em}
\sffamily
\begin{tabular}{m{4.5cm} p{13.5cm} }

\includegraphics{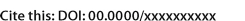} & \noindent\LARGE{\textbf{In-Depth Exploration of Catalytic Sites on Amorphous Solid Water:  I. The Astrosynthesis of Aminomethanol$^\dag$}} \\
\vspace{0.3cm} & \vspace{0.3cm} \\

 & \noindent\large{Giulia M. Bovolenta\textit{$^{a,b}$}, Gabriela Silva-Vera\textit{$^{a}$}, Stefano Bovino\textit{$^{c,d,e}$}, German Molpeceres\textit{$^{f}$}, Johannes Kästner\textit{$^{g}$} and Stefan Vogt-Geisse$^{\ast}$\textit{$^{a}$}} \\

\includegraphics{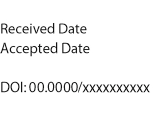} & \noindent\normalsize{

Chemical processes taking place on ice-grain mantles are pivotal to the complex chemistry of interstellar environments.
In this study, we conducted a comprehensive analysis of the catalytic 
effects of an amorphous solid water (ASW) surface on the reaction between 
ammonia (\ce{NH3}) and formaldehyde (\ce{H2CO}) to form aminomethanol
(\ce{NH2CH2OH}) using density functional theory.  
We identified potential catalytic sites based on the binding energy 
distribution of \ce{NH3} and \ce{H2CO} reactants,  on a set-of-clusters surface 
model composed of 22 water molecules and found a total of 14
reaction paths. Our results indicate that the catalytic sites 
can be categorized into four groups, depending on the interactions of the 
carbonyl oxygen and the amino group with the ice surface in the reactant complex.
A detailed analysis of the reaction mechanism using Intrinsic Reaction Coordinate and reaction force 
analysis, revealed three distinct chemical events for this reaction: formation
of the C--N bond, breaking of the N--H bond, and formation of the O--H hydroxyl 
bond. Depending on the type of catalytic site, these events can occur within a single,
concerted, albeit asynchronous,  step,  or can be isolated in a step-wise mechanism, with the lowest overall 
transition state energy observed at 1.3 kcal $\mathrm{mol}^{-1}$.
A key requirement for the low-energy mechanism is the presence 
of a pair of dangling OH bonds on the surface, found at 5\% of the potential catalytic sites 
on an ASW porous surface.

} \\

\end{tabular}

 \end{@twocolumnfalse} \vspace{0.6cm}

  ]

\renewcommand*\rmdefault{bch}\normalfont\upshape
\rmfamily
\section*{}
\vspace{-1cm}

\footnotetext{\textit{$^{a}$~Departamento de Físico-Química, Facultad de Ciencias Químicas, Universidad de Concepción, Concepción, Chile}}
\footnotetext{\textit{$^{b}$~Atomistic Simulations, Italian Institute of Technology, 16152 Genova, Italy}}
\footnotetext{\textit{$^{c}$~Chemistry Department, Sapienza University of Rome, P.le A. Moro, 00185 Rome, Italy}}
\footnotetext{\textit{$^d$~INAF - Osservatorio Astrofisico di Arcetri, Largo E. Fermi 5, 50125 Firenze, Italy}}
\footnotetext{\textit{$^e$~ Departamento de Astronomía, Facultad de Ciencias Físicas y Matemáticas, Universidad de Concepción, Av. Esteban Iturra s/n Barrio Universitario, Concepción, Chile}}
\footnotetext{\textit{$^{f}$~Departamento de Astrofísica Molecular
Instituto de Física Fundamental (IFF-CSIC), Madrid, Spain }}
\footnotetext{\textit{$^{g}$~Institute for Theoretical Chemistry, University of Stuttgart, Pfaffenwaldring 55, 70569 Stuttgart, Germany }}
\footnotetext{\dag~Electronic Supplementary Information (ESI) available}



\section{Introduction}
In molecular clouds of the interstellar medium,  where temperatures are 
extremely low (<20 K), interstellar dust grains are covered 
by thick ice mantles, which are significant reservoirs of 
chemical species.  Recently, thanks to the extraordinary 
sensitivity offered by the James Webb Space Telescope (JWST), 
it has been possible to carry out ice observations with unprecedented resolution\cite{mcclure_ice_2023}. In terms of composition\cite{tielens_interstellar_1998,boogert_observations_2015},
water is known to be the primary ice component, along with \ce{CO}, \ce{NH3},
\ce{CO2} and \ce{CH4}.  Although precise information about ice morphology 
is still lacking, water is assumed to be in the so-called amorphous solid water phase (ASW), 
and to feature a certain porosity\cite{palumbo_formation_2006}.
Experiments on ASW 
show\cite{amiaud_interaction_2006,watanabe_direct_2010, he_binding_2016}
that, due to the amorphous nature of the surface, 
molecules interact with ice in different ways  depending on the 
type of sites.  The strength of the interaction  can be described by the 
binding energy (BE) of the molecule with respect to the ice surface.
Therefore, every surface chemical process may depend on the adsorption profile of the 
participant molecules, which is characterized by the distribution of binding energies, 
although a specific connection between binding energies and catalytic activity is still missing.
According to experimental \cite{he_interaction_2011, noble_thermal_2012}
and theoretical\cite{bovolenta_high_2020} findings, these BE 
distributions resemble a Gaussian function.  In a novel \textit{multi-binding} approach, 
the single BE value for a certain species can be replaced with a BE Gaussian distribution, 
as it has been recently implemented\cite{grassi_novel_2020} in astrochemical kinetic models. 

Few theoretical studies have explored the impact of the surface morphology on interstellar 
reactive processes. \citet{song_formation_2016} studied the hydrogenation 
of \ce{HCNO} using QM/MM techniques on a hemispherical water surface with a radius of 
34 \text{\AA} and obtained 81 binding sites. They employed BHLYP-D3/def2-TZVPD in the QM region 
on five different \ce{H2CO} adsorption sites with different BE and found no 
significant impact on the transition state energies. In a more recent work, 
\citet{paiva_glycolaldehyde_2023} studied the addition reaction of \ce{H2CO} and \ce{HCO}
on 10 different periodic slabs, containing 25 water molecules randomly arranged 
within the unit cell. They found a significant catalytic effect of the water surface with respect to 
the gas phase reflected by a 50 \% reduction in the reaction barrier.

A good target for applying the \textit{multi-binding} approach, to 
evaluate the catalytic effect of the plethora of possible binding sites 
present on an ASW surface, is the first step of the Strecker synthesis. 
This series of  reactions  has enjoyed a longstanding fascination in the
astrochemical community, as it provides a pathway to the formation of glycine, the 
simplest amino acid, from  relatively abundant species in the
interstellar medium: \ce{NH3}, \ce{H2CO} and \ce{HCN}. The first step  
corresponds to the nucleophilic addition of \ce{NH3} 
to \ce{H2CO} to  form \ce{NH2CH2OH} (aminomethanol, AMeOH). 

\begin{equation}
\label{eq:S1}
 \ce{H2CO} + \ce{NH3} \rightarrow  \ce{NH2CH2OH} 
\end{equation}

Laboratory preparation and isolation of AMeOH have been elusive. The species has been tentatively 
formed in a seminal experimental work \cite{schutte_experimental_1993} using ice-analogs,  
using a 10 K mixture of \ce{H2O}:\ce{NH3}:\ce{H2CO}, co-deposited on a cold surface 
and subsequently warmed up, but the identification of the reaction product was considered ambiguous. 
The experiment has been repeated by  \citet{bossa_nh_2009}, 
confirming the previous findings and estimating AMeOH formation to take place between 80 and 100 K.  
Furthermore, AMeOH has also been identified in a temperature-programmed desorption experiment,  starting from  \ce{CH3NH2} and \ce{O2} ices,  
upon exposure to energetic electrons\cite{singh_experimental_2022}, showing unexpected kinetic 
stability under extreme environments. Nevertheless, AMeOH has hitherto not been observed in 
the interstellar medium.

The formation of AMeOH via the first step of the Strecker synthesis has also been 
studied theoretically.  The gas phase barrier of this reaction has been 
determined to be high for interstellar conditions, amounting to 34 kcal $\mathrm{mol}^{-1}$\cite{woon_ab_1999}.
The water-catalyzed reaction has been studied in the presence of small 
water clusters \cite{woon_ab_1999, courmier_computational_2005, chen_theoretical_2011} 
and with the addition of implicit solvation using a polarizable continuum model.
The incorporation of the water cluster results in a significant lowering 
of the reaction barrier by 20 kcal $\mathrm{mol}^{-1}$.  To date, the most accurate barrier  
for the reaction in the presence of a water dimer, was provided by
\citet{courmier_computational_2005} at CCSD(T)/6-311+G**//MP2/6-311+G** level of theory, 
amounting to 14 kcal $\mathrm{mol}^{-1}$. 
Finally, the same reaction has been studied using larger cluster models. \citet{rimola_deep-space_2010}
simulated the reaction on a 18-molecules crystalline water ice surface model at 
DFT/B3LYP level,  with a reported TS energy of 9.6 kcal $\mathrm{mol}^{-1}$.

In this work, we present a detailed analysis of the surface catalytic effect by 
studying the addition reaction of \ce{NH3} and \ce{H2CO} on different catalytic 
sites of an ASW surface, spanned by a set 
of amorphized 22-water-molecules clusters. We also study the same reaction inside a 
nano-pore, derived from a periodic ASW surface of  500 water molecules. We explore the 
different reaction pathways by means of DFT and analyze the reaction mechanism using Intrinsic Reaction Coordinate (IRC)
energy profiles, Natural Bond Orbital (NBO) bond order derivatives, and reaction force analysis. 
In Sec. \ref{sec:theo}, we describe the aspects of the methodology used in this work. In Sec. \ref{sec:be_react}, we 
present the BE profiles of \ce{NH3} and \ce{H2CO}, followed by the reaction path and Potential Energy Surface (PES) 
analysis of the reaction in different binding scenarios, on the ASW set of clusters (Sec. \ref{sec:gen_aspS1}, \ref{sec:resultsASW}), and in a nano-pore model (Sec. \ref{sec:S1_cavity}). Finally, we discuss the results and 
the astrophysical implications in Sec. \ref{sec:discussion}. 

\section{Theoretical Methods}\label{sec:theo}
\subsection{DFT model chemistry}\label{sec:ref} 
 We performed an extensive benchmark of DFT methods. The reference system for Reaction \ref{eq:S1}
 is constituted by the reactants coordinated to two water molecules 
 ($\ce{NH3} + \ce{H2CO} + 2\ce{H2O}$), acting as proton transfer intermediaries during the reaction,
in an arrangement commonly named as \textit{proton relay}.
The reference system geometry is DF-CCSD(T)-F12/cc-pVDZ-F12\cite{gyorffy_analytical_2018,sylvetsky_aug-cc-pvnz-f12_2017}, which has been shown to provide excellent 
geometries\cite{warden_efficient_2020}. The energies of stationary points have been computed 
at CCSD(T)\cite{bozkaya_analytic_2017} using a complete basis set (CBS)  extrapolation\cite{helgaker_basis-set_1997,DunningT.H.2001}. For the 
CBS extraploation we employed the aug-cc-pVXZ with X=D,T,Q for SCF energies, X=T,Q for MP2 correlation  energies and X=D,T for CCSD and CCSD(T) correlation energies extrapolation using the extrapolation routines implemented
in the Psi4 driver\cite{parrish_psi4_2017}. We took into account around 53 DFT functionals
for the geometry benchmark belonging to different classes, and  two different  basis sets: def2-SVP and 
def2-TZVP\cite{weigend2005a}. This is due to the fact that a double $\zeta$ basis is used to study the reaction 
on a variety of binding sites on the larger surfaces, hence the need to assess the consistency of 
a specific DFT functional with the two tiers of method and basis. 
Dispersion effects are treated using D3BJ\cite{grimme_consistent_2010,grimme_effect_2011} and D4\cite{caldeweyher_generally_2019} correction. The geometry benchmark has been carried out using \textsc{Orca}\cite{neese_orca_2020} software, adopting TIGHTOPT thresholds criteria.
The best method for the lower tier equilibrium geometry is BHandHLYP-D4/def2-SVP\cite{becke_new_1993},  
with an average Root Mean Square Deviation (RMSD) error of 0.1 \text{\AA} with respect to the CCSD(T) reference.
Regarding the energy benchmark, 258 DFT functionals are considered. The best ones are 
BMK/def2-TZVP\cite{boese_development_2004} and $\omega-$B97M-D3BJ/def2-TZVP (a modified version of $\omega-$B97M-V augmented with D3BJ correction\cite{najibi_nonlocal_2018})  
that show a mean absolute error (MAE)  below 1 kcal $\mathrm{mol^{-1}}$.   

To summarize,
the benchmark allows to identify two suitable methods: Tier 1: $\omega$-B97M-D3BJ/def2-TZVP // BHandHLYP-D4/def2-SVP and Tier 2: BMK/def2-TZVP // BHandHLYP-D3BJ/def2-SVP. We used the former with the \textsc{Orca} software, and  the latter with \textsc{Gaussian}\cite{g16} software, due to the unavailability of Tier 1. For all calculations of the reactive sites on the ASW clusters we used the default optimization thresholds. 
Full benchmark results can be found in the Supplementary Material$\dag$  (Sec. S1). 
High level CCSD(T)-F12 optimizations are performed using \textsc{Molpro}\cite{werner_molpro_2020}.

\subsection{ASW Modeling}
\subsubsection*{Set of clusters}
The ice surfaces used for adsorption and reactivity studies are modeled according to the 
\textit{cluster approach}\cite{shimonishi_adsorption_2018}. We used homogeneous amorphous water clusters of 22 water molecules each (Fig. \ref{fig:asw_models_clusters}), modeled by means of \textit{ab initio} molecular dynamics (AIMD).   
The cluster size is selected in order to guarantee a reasonable number of available 
sites, while at the same time being able to use high-level model 
chemistry.  The steps to obtain the ASW models  have been illustrated in a previous work
\cite{bovolenta_high_2020}. Amorphization AIMD simulations on the cluster models are performed at BLYP/def2-SVP\cite{becke_density-functional_1988,lee_development_1988} 
level, adding D3 Grimme  \cite{grimme_consistent_2010} 
correction for dispersion 
interactions, as implemented in
\textsc{Terachem}.  
\cite{ufimtsev_quantum_2009, titov_generating_2013}.

\subsubsection*{Nano-pore model}\label{sec:cavity_ice}

\begin{figure}

\centering
\begin{subfigure}[b]{0.5\textwidth}
\includegraphics[width=\textwidth]{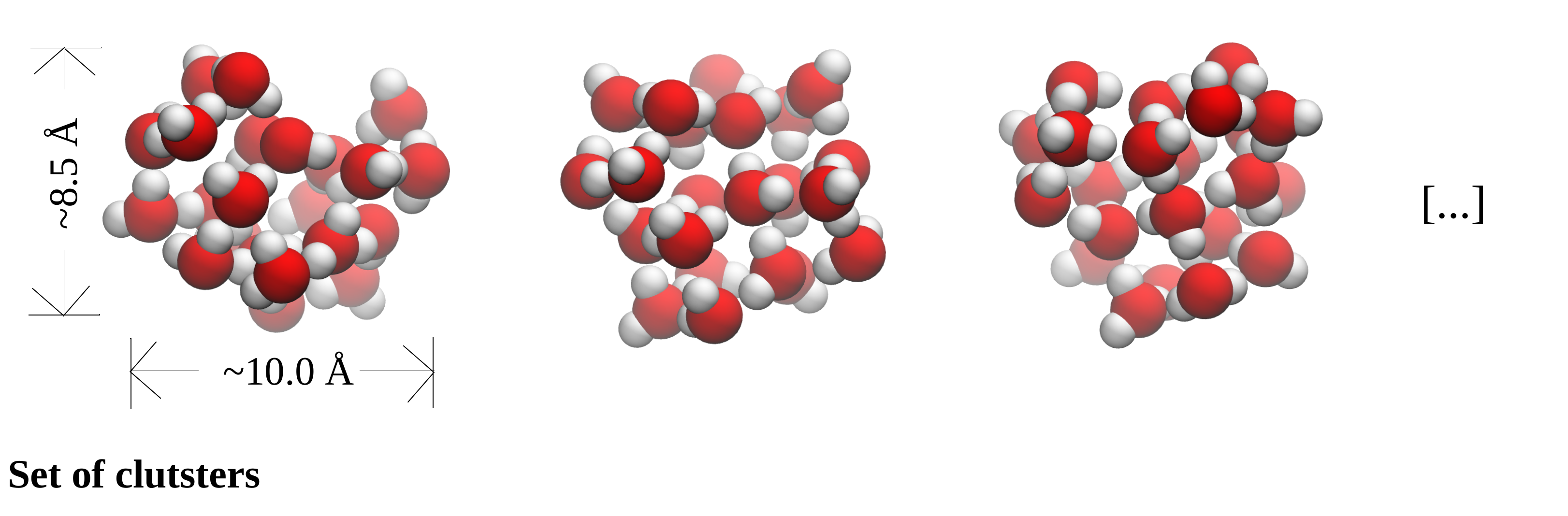}
   \caption{}
   \label{fig:asw_models_clusters}
\end{subfigure}
\vspace{0.2cm}

\begin{subfigure}[b]{0.5\textwidth}\includegraphics[width=0.98\textwidth]{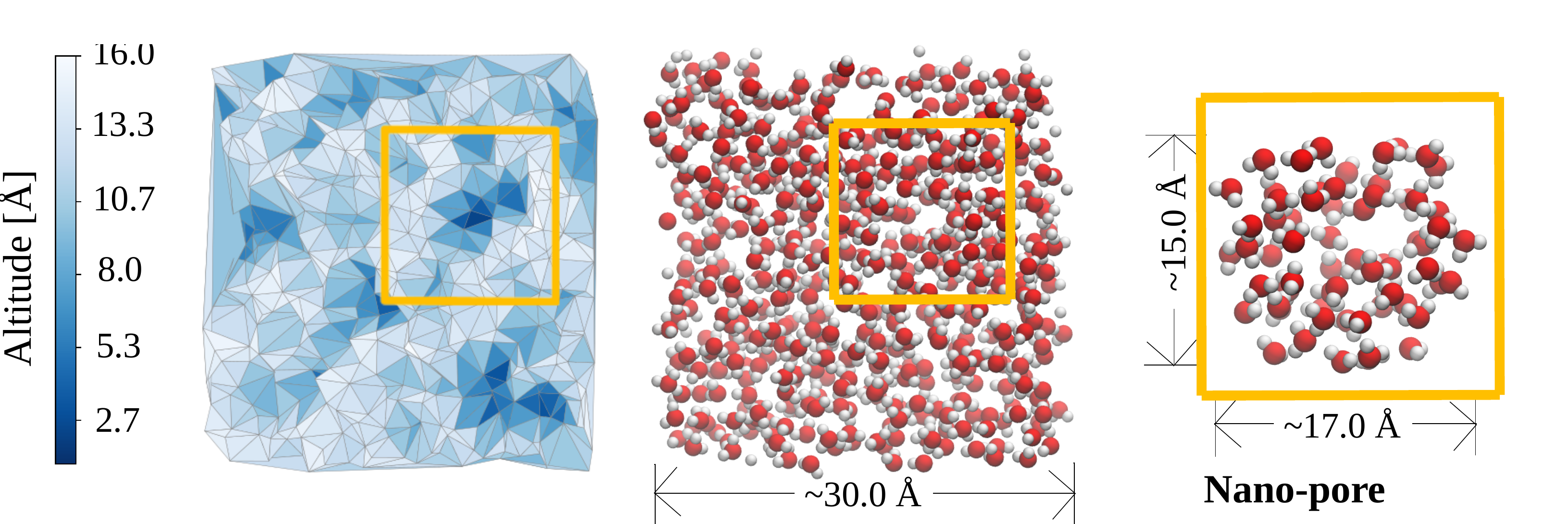}
   \caption{}
   \label{fig:asw_models}
\end{subfigure}
\caption{a) Some of the 20 homogeneous 22-water-molecules amorphous clusters used in this work for BE evaluation and
reactivity studies. After modeling through AIMD techniques, the structures undergo
geometry optimization. b) Generation of the nano-pore model. From the left: altitude map (see Supplementary Material$\dag$ Sec. S6.2) of an ASW slab composed of 500 water molecules, with an initial density of 0.8 g $\mathrm{cm^{-3}}$ and top view of the corresponding periodic cell. One of the suitable nano-porous regions of the slab is highlighted in yellow. To the right:  the spherical pore site as extracted from the slab, prior to geometry optimization.
The color scheme for the atoms is red for O and white for H.}

\end{figure}
 
ASW in interstellar environments is thought to be partly porous\cite{hama_surface_2013}. Such pores sites facilitate chemical encounters of species adsorbed on the ice and might enhance surface reactivity.  
In order to generate a nano-pore, we used a periodic surface slab with low initial density (0.8 g $\mathrm{cm^{-3}}$), as a starting point.
The periodic slab of 500 water molecules has been modeled 
using a \textit{ad hoc} trained Machine Learned Potential (MLP)\cite{zaverkin_gaussian_2020,zaverkin_fast_2021,zaverkin_neural-network_2022},
that will be employed on a subsequent work. Details of the slab generation and the MLP training and validation can be found in the Supplementary Material$\dag$, Sec. S5.
One of the periodic surfaces is reported in Fig.  \ref{fig:asw_models}. 
In order to visually appreciate the presence of concave regions ('valleys'), pores, and 'crests' on the surface, the figure includes an altitude map (see also Supplementary Material$\dag$ Sec. S6.2 for details about Tri-Surface plots generation).
We extracted a spherical portion of one of the pores sites of nanometric size 
from the slab (highlighted in Fig. \ref{fig:asw_models})
consisting of 64 water molecules,  
constraining the cartesian positions of the atoms of the outer sphere, in order to preserve the pore during geometry optimization.\\
To run MD simulations on the periodic systems,  the GMNN program\cite{zaverkin_gaussian_2020} is interfaced to the ASE package\cite{hjorth_larsen_atomic_2017}.
\textsc{Orca}\cite{neese_orca_2020} software is used to compute DFT energies and gradients that constitute MLP training set.

\subsection{Binding Energy, Binding Energy Distributions, Interaction Energy}
The BE of a species ($i$) adsorbed on a surface ($ice$) is defined as:    
\begin{equation}
BE(i) = E_{sup} - (E_{ice}+ E_i)
\label{eq:BE}
\end{equation}
where $E_{sup}$ stands for the energy of the supermolecule formed by the adsorbate bound to the surface, $E_{ice}$ refers to the water cluster energy, and $E_i$ is the energy of the adsorbate.
The BE is assumed to be a positive quantity, according to convention. 
Using a BE distribution (BEd) of values reflects 
a more realistic desorption behavior for molecules adsorbed on ASW ice. 
All the BEd in this work have been produced  
using BEEP computational platform and protocol\cite{bovolenta_binding_2022}. The equilibrium geometry is of HF-3c/MINIX\cite{sure_corrected_2013} quality, as it is a cost-effective method for these systems, while the level of theory used for the energy has been chosen according to a DFT benchmark\cite{bovolenta_binding_2022}. 
According to the BEd of a species, it is possible to define different intervals of BE. For a certain species $i$:
\begin{itemize} 
\item Low-BE($i$):  
BE < 0.2$\times$mean(BEd)  
\item High-BE($i$): BE > 0.7$\times$mean(BEd)
\end{itemize}

The BE can be further decomposed into two components: the deformation energy $DE(i)$ and interaction energy $IE(i)$, such that $BE(i) = DE(i) - IE(i)$.
We computed the interaction energy using a zeroth order Symmetry Adapted Perturbation Theory (SAPT0) analysis\cite{jeziorski_perturbation_1994}, together with a jun-cc-pVDZ\cite{papajak_perspectives_2011} basis set. This allowed us to directly quantify the strength of the non-covalent intermolecular interaction between the reactant complex (R) and the ASW surface (W) in the bound configuration:
\begin{equation}
IE(R) = - IE(\ce{NH3} + \ce{H2CO} \dotsb \ce{W})
\label{eq:ie_r}
\end{equation}

Binding sites optimization and BE computations are performed using \textsc{Psi4} \cite{parrish_psi4_2017} \textit{via} the platform BEEP\cite{bovolenta_binding_2022,smith_molssi_2020}. All SAPT0 computations use the density-fitted implementation\cite{hohenstein_density_2010,hohenstein_large-scale_2011} provided in \textsc{Psi4}.

\subsection{Reactive sites sampling}\label{sec:site_samp}
In agreement with the 
\textit{multi-binding} framework, the reactions under study have been carried out on a variety of sites on
the ASW surfaces.  The procedure to obtain the set of transition states
for Reaction \ref{eq:S1} has been the following:
\begin{enumerate}
\item Selection of several suitable Low-BE and High-BE binding sites for each reactant. The selection is based on the binding mode analysis 
which allows to identify the arrangement for the reaction to take place (\textit{vide infra}, Sec. \ref{sec:be_react})
\item Extensive sampling of the first fragment around the other in Low-BE and High-BE sites, and vice versa. 
This step provides a set of potential reaction sites.
\item Transition state (TS) search and characterization 
for the whole set, exploring the PES by means of relaxed scan and nudged elastic band methods\cite{jonsson_nudged_1998}.\\
The minimum energy structures are identified and optimized using \textsc{Orca}. 

\end{enumerate}
\subsection{Reaction Force analysis}  
The most widespread method to locate a reaction path is  
tracing the IRC\cite{fukui_path_1981}, which corresponds to the minimum energy path in
mass-weighted coordinates. Useful information about the mechanism of a chemical process
can be obtained from the reaction force profile. For a certain process, 
the potential energy $E(\xi)$ of the system along the intrinsic reaction coordinate ($\xi$) has an associated 
reaction force $F(\xi)$, deﬁned by:

\begin{equation}
\label{eq:reaction_force}
    F(\xi) = -\frac{dE}{d\xi}
\end{equation}
It has been shown\cite{toro-labbe_characterization_1999} that the critical points of $F(\xi)$ define regions along $\xi$
in which different reactive events take place. The reactive events are identified as 
inflection points of the reaction force profile. Within each reaction event $i$, it is 
possible to identify three regions, limited within the critical points present in $F(\xi)$, 
namely the local minimum, at $\xi_{min,i}$,  and the local maximum, at $\xi_{max,i}$.  The pre- and post-
event region  ($\xi \leq \xi_{min}$, $\xi_{max} \leq \xi$) are characterized by
structural preparation and relaxation of the participating species, respectively. On the other hand, the event region itself is governed by changes in the electron density associated with bond formation and dissociation processes.

In general, a single kinetic step of a  chemical reaction can be composed of $N$ different reaction events, such that the total reaction energy corresponds to the sum of the energy of the individual reactive events:

\begin{equation}
    \label{eq:work}
    \Delta E^{o} = \sum_{i=1}^{N} \Delta E_{i}
\end{equation}

\noindent The event energy $\Delta E_{i}$ will be positive if the event takes place before the TS 
and negative if it takes place past the TS.
Conversely, the TS energy barrier ($\Delta E^{\ddagger}$) is partitioned among the events leading up to the 
TS, for a given reaction:

\begin{equation}
    \label{eq:work}
    \Delta E^{\ddagger} = \int_{\xi_{0}}^{\xi_K} F(\xi)d\xi 
 =  \sum_{i=1}^K \Delta E^{\ddagger}_i
\end{equation}
where $K$ is the number of events that take place before reaching the TS, $\xi_0 $ and  $\xi_K$ corresponds to the reaction coordinate of the reactants and 
TS respectively. 
In turn, the energy required for a certain event $i$ can be obtained by integrating the reaction 
force profile:

\begin{equation}
    \label{eq:work}
    \Delta E^{\ddagger}_{i} =  \int_{\xi_{i-1}}^{\xi_i} F(\xi)d\xi 
\end{equation}
Events leading up to the TS are denominated hidden transition states (h-TS)\cite{kraka_computational_2010}, which are associated with hidden intermediates (h-I). Such intermediates are not observable along the reaction energy profile and, hence, can only be estimated to lie between the h-TS and TS.  
The integration of the force for the reaction events intervals  allows
to estimate the energy expenditure associated to 
each phase of the chemical process and proved to be a 
valuable partition for quantifying the isolated chemical changes.\\
IRC calculations are performed with \textsc{Gaussian}\cite{g16}
after re-optimization of the stationary states with that same software. Analysis of the IRC profiles is carried out using Kudi\cite{Vogt-Geisse2016}.
Natural Atomic Orbital(NAO) population analysis uses the NBO software version\cite{glendening_nbo_2013} implemented in \textsc{Gaussian} and \textsc{Orca}.  

\section{Results}
\begin{figure}[t]
    \centering
    \includegraphics[width=0.48\textwidth]{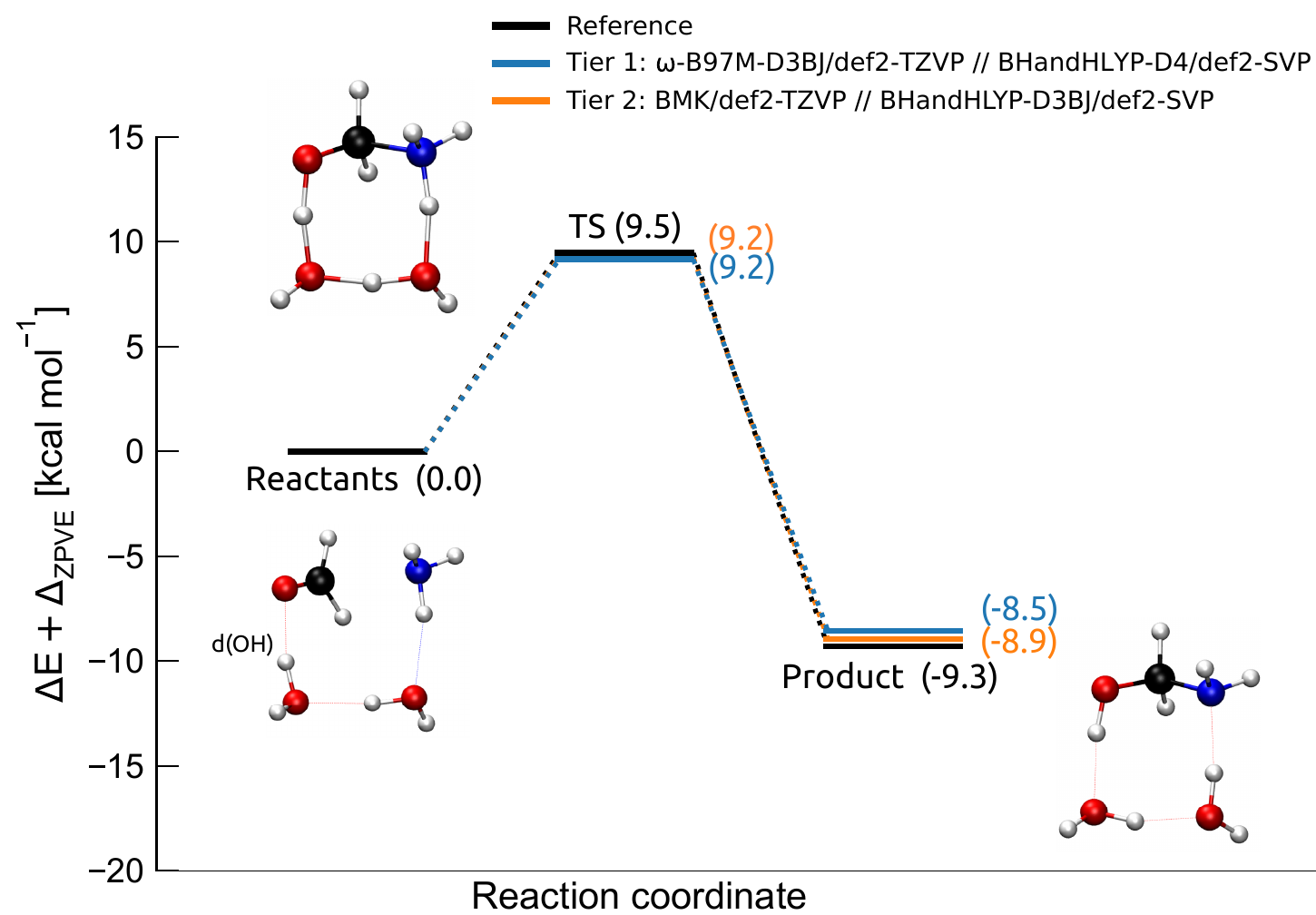}
	\caption{Energy diagram for Reaction \ref{eq:S1}, reference system (\ce{NH3} + \ce{H2CO} + \ce{2H2O}), using CCSD(T)/CBS // DF-CCSD(T)-F12/cc-pVDZ-F12 level of theory  (solid black line), $\omega$-B97M-D3BJ/def2-TZVP // BHandHLYP-D4/def2-SVP (blue solid line, Tier 1) and BMK/def2-TZVP // BHandHLYP-D3BJ/def2-SVP (orange solid line, Tier 2)  levels of theory.  Water-OH 
groups not engaged 
in any H-bond with other water molecules are labelled d(OH) (\textit{dangling-OH} bonds).
 The color scheme for the atoms is red for O, black for C, blue for N, and white for H.}
     \label{fig:S1_ref}
\end{figure}

Fig. \ref{fig:S1_ref} shows the reactants, transition state, and product of Reaction \ref{eq:S1}, studied 
on a two-water molecules cluster using high-level model chemistry (see Sec. \ref{sec:ref}). Overall, two new bonds are formed (C--N and O--H)
and one bond is broken (N--H) to yield the AMeOH product. The reactive 
events involving the proton transfer are assisted by the two water molecules 
that relay the proton from the amino end to form the hydroxyl moiety in the 
amino alcohol. On this minimal water surface model, the reaction is exothermic 
by -9.3 kcal $\mathrm{mol^{-1}}$  with a TS barrier of 9.5 kcal $\mathrm{mol^{-1}}$.
Fig. \ref{fig:S1_ref} also reports the two Tiers corresponding to the best performing methods of 
the TS and reaction energy benchmark, which shows an excellent agreement between the DFT model chemistry and the 
\textit{ab initio} reference (see Sec. \ref{sec:theo}).
\subsection{Binding modes and energies of \ce{NH3} and \ce{H2CO}}\label{sec:be_react}
With the goal of analyzing the binding patterns of \ce{NH3} 
and \ce{H2CO} to the ASW surface that
enable the formation of AMeOH, we analyzed the binding modes within 
the BE distribution that we obtained from the BEEP database, 
on a set of 22-water-molecules clusters. For Reaction \ref{eq:S1} to take place, 
\ce{NH3} has to be acting as a hydrogen-bond (H-bond) donor with respect to the ice, 
while carbonyl-O  of \ce{H2CO} is placed as H-bond 
acceptor. The surface molecule acting as a donor to the carbonyl-O presents an OH 
group pointing upwards away from the surface (not engaged 
in any other H-bond). Such surface OH bonds are labeled \textit{dangling-OH} bonds - d(OH).

\begin{figure}[t]
    \centering
    \includegraphics[width=0.48\textwidth]{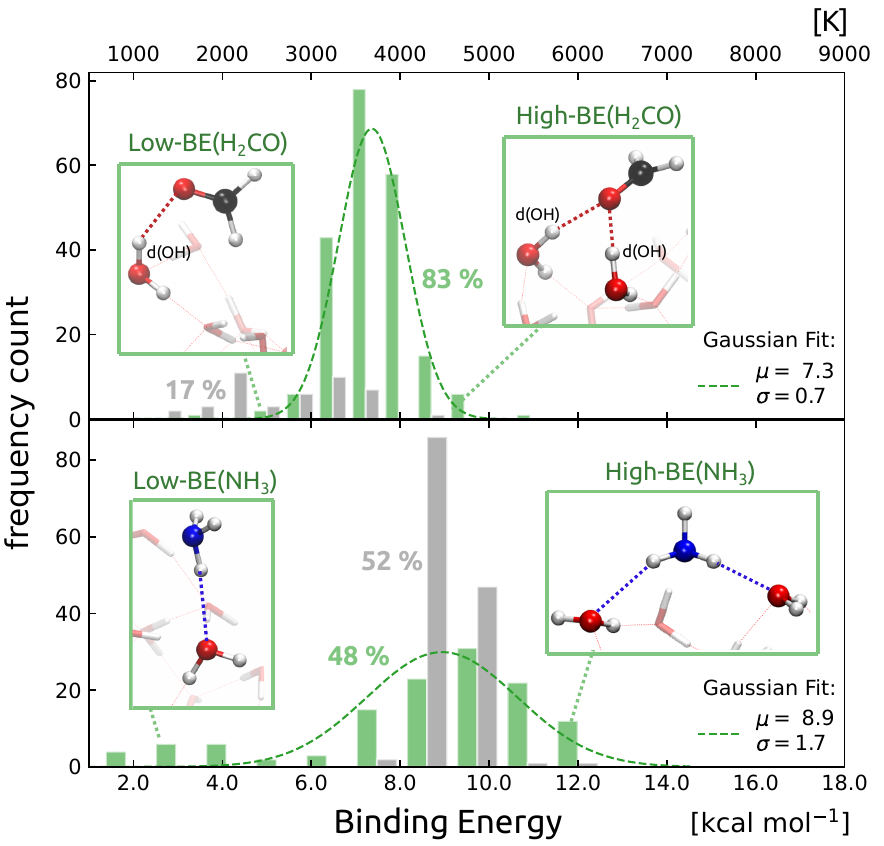}
	\caption{Histogram of the BE distribution of \ce{H2CO} (upper panel) and \ce{NH3} (lower panel)
     computed on a set-of-clusters model with a cluster size of 22 water molecules, using HF-3c/MINIX geometries. According to the benchmark results, the energy has been computed at $\omega$−PBE/def2-TZVP level of theory for \ce{NH3} and CAM-B3LYP/def2-TZVP level of theory for \ce{H2CO}.  The ZPVE correction has not been included. 
     BE values are given in K (upper scale) and kcal $\mathrm{mol^{-1}}$ (lower scale).  The green color corresponds to the binding mode that is conducive to the formation of AMeOH through Reaction \ref{eq:S1},  while binding modes that could not directly engage in a reactive encounter are in grey. Mean BE ($\mu$) 
     and standard deviation ($\sigma$) of a Gaussian fit of the distribution are reported in kcal $\mathrm{mol^{-1}}$  for the suitable binding mode (green). The fit is  obtained using a bootstrap method, see Supplementary Material$\dag$ Sec. S6.1.
     The insets show an example of  High-BE and Low-BE molecules in a favorable orientation for the reaction. Water-OH 
groups pointing upwards away from the surface (not engaged 
in any other H-bond) are labelled d(OH) (\textit{dangling-OH} bonds). The color scheme for the atoms is red for O, black for C, blue for N, and white for H.}
     \label{fig:be_modes_S1}
\end{figure}

Fig. \ref{fig:be_modes_S1} reports the BE distributions for both molecules. 
The fraction of binding sites compatible with Reaction \ref{eq:S1} has been 
highlighted in green, and represents the 48 \% and 83 \%  
for \ce{NH3} and 
\ce{H2CO}, respectively. 
Among the binding sites, we identified different binding regimes: Low-BE sites of \ce{NH3} and \ce{H2CO}, 
in which there is only one ice H-bond acceptor and donor group, respectively, and
High-BE sites, where there are two water molecules coordinated to each fragment. The latter corresponds to 
the interaction with two acceptor groups - High-BE(\ce{NH3})- and to
two d(OH) bonds - High-BE(\ce{H2CO}). 
Combining those possible 
binding regimes results in 4 limiting patterns for the reactants in the bound configuration: \begin{itemize}
    \item Low-BE(\ce{NH3)}/Low-BE(\ce{H2CO})
    \item High-BE(\ce{NH3)}/Low-BE(\ce{H2CO})
    \item Low-BE(\ce{NH3)}/High-BE(\ce{H2CO})
    \item High-BE(\ce{NH3)}/High-BE(\ce{H2CO})
\end{itemize}

\begin{figure}[t]
    \centering
\includegraphics[width=0.3\textwidth]{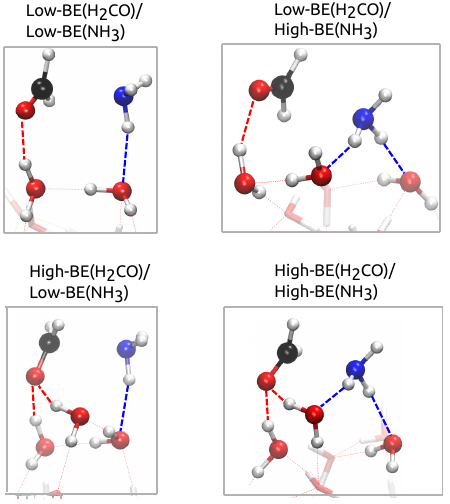}
	\caption{Example of structures belonging to the four groups of reactive sites identified for Reaction \ref{eq:S1}, using BHandDHLYP-D4/def2-SVP geometries. 
     The color scheme for the atoms is red for O, black for C, blue for N, and white for H. H-bonds established by the reactants have been highlighted and colored accordingly.}
     \label{fig:R_S1}
\end{figure} 

\subsection{Binding site sampling}\label{sec:gen_aspS1}

We sampled selected binding sites representing the different binding regimes 
shown in the previous section, with its complementary reactant molecule 
for Reaction \ref{eq:S1}, following the procedure illustrated in Sec. \ref{sec:site_samp}. In total, we identified 14 reaction-conducing catalytic sites on different ASW clusters, traceable to 
the aforementioned 4 categories. In order to further characterize 
and quantify the strength of the interaction of the reactant complex, we computed the $IE(R)$ 
with respect to the ASW surface (Eq. \ref{eq:ie_r}), for 
the different reactive sites. The results are reported in Table \ref{tab:IE}.  
There is a good agreement between the binding regime of the individual reactants and the total $IE(R)$, as the reactant complexes that correspond to High-BE orientations display the highest $IE(R)$, while the ones that are categorized in the Low-BE regime also have the lowest average $IE(R)$.

\begin{table}[h]
    \centering   
\caption{Interaction energy between the reactants and the ASW surface ($IE(R)$), calculated at SAPT0 level with jun-cc-pVDZ basis, according to Eq.\ref{eq:ie_r}. Average (Avg) and standard deviation (Std) for each binding regime are reported, as well. Values in kcal $\mathrm{mol^{-1}}$. The listed reactant states are shown in Fig. S6, Supplementary Material$\dagger$.} 
  \begin{tabular}{l l c}
  \hline
 & $IE(R)$         & Avg (Std)  \\
                  \hline
\textit{ASW clusters:} & & \\
Low-BE(\ce{NH3)}/\/Low-BE(\ce{H2CO})  & \\

     &     A     11.7 & \\
     &     B     10.5 & \\
     &     C      11.3&  \\
     &     D     11.6 &  \\
     &     E     8.0  &  \\
     &    & 10.6 (1.4)   \\
   High-BE(\ce{NH3)}/\/Low-BE(\ce{H2CO}) & \\
     &    A     13.7 &   \\
     &    B     11.3 & \\
     &    C     12.2 &\\
     &           & 12.4(1.0)\\  
Low-BE(\ce{NH3)}/\/High-BE(\ce{H2CO}) & \\
    &    A      16.6 & \\
    &    B      16.6 &\\
    &    C        17.6 & \\
    &    D      18.9& \\
    &  &   17.7(0.9)\\
High-BE(\ce{NH3)}/\/High-BE(\ce{H2CO}) &  \\
 & A  21.8 \\
             & B 21.7\\
 &  & 21.7(0.05) \\\\

 \textit{Porous ASW:} & A 32.3 & \\
 
\end{tabular}
\label{tab:IE}
\end{table}

\subsection{AMeOH formation on ASW clusters}\label{sec:resultsASW}
\subsubsection{Low-BE(\ce{NH3)}/Low-BE(\ce{H2CO})}
\begin{figure*}
    \centering
    \includegraphics[width=0.9\textwidth]{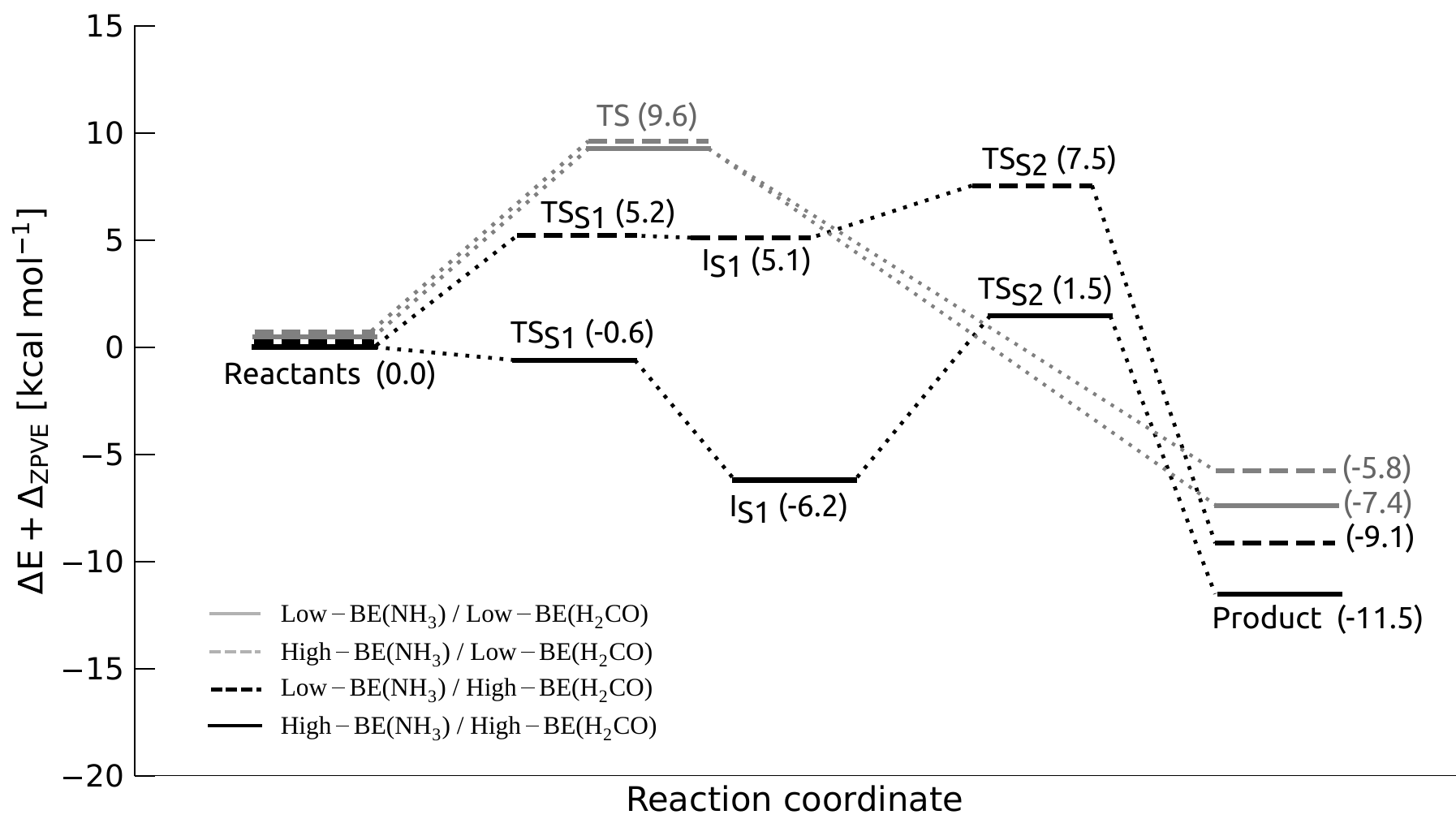}
     \caption{Energy diagram for the lowest energy pathways of each binding regime (system A, see Table \ref{tab:deltaE_S1}), using BHandHLYP-D4/def2-SVP geometries and  $\omega$-B97M-D3BJ/def2-TZVP energies. 
     Note that in the High-BE(\ce{NH3})/High-BE(\ce{H2CO}) regime, \ce{TS_{S1}} has a negative value after ZPVE correction. Without the correction the value is 0.1 kcal $\mathrm{mol^{-1}}$, see Table S3 in Supplementary Material$ \dagger $.}
      \label{fig:S1_all_energy}
\end{figure*}

The first case, where both fragments are loosely bound to the surface,  
represents a  situation where the effect of the ice  
on the reactive site resembles  
the water dimer model system: both the reactants establish one H-bond each with the surface,
as they interact solely with the water molecules that will assist the 
proton transfer.  
Five structures correspond to this case. 
Estimation of the \textit{IE(R)} between the reactants and the water surface,  
indicates that 
those structures presents the smallest average \textit{IE(R)} 
value (10.6 kcal $\mathrm{mol^{-1}}$) of all the cases (see Table \ref{tab:IE}).\\  
For this case, Reaction \ref{eq:S1}  
is concerted, and one TS was located.
Table \ref{tab:deltaE_S1}, reports the
TS energy for the structures in this group (systems A-E), while the diagram in Fig.  \ref{fig:S1_all_energy},  shows only the lowest energy value (system A, solid grey line). 
The average TS barrier of 10.9 kcal $\mathrm{mol^{-1}}$   
is in line with the
model system (Fig. \ref{fig:S1_ref}), along with the small variation of the average TS barrier, when the proton involves two water molecules (cases A--D).  It is 
worth noting that a proton relay mechanism with a single water 
molecule, as in case E,  results in a significantly 
higher TS energy and a more exothermic reaction.

\begin{figure*}[ht!]
    \centering
    \includegraphics[width=\textwidth]{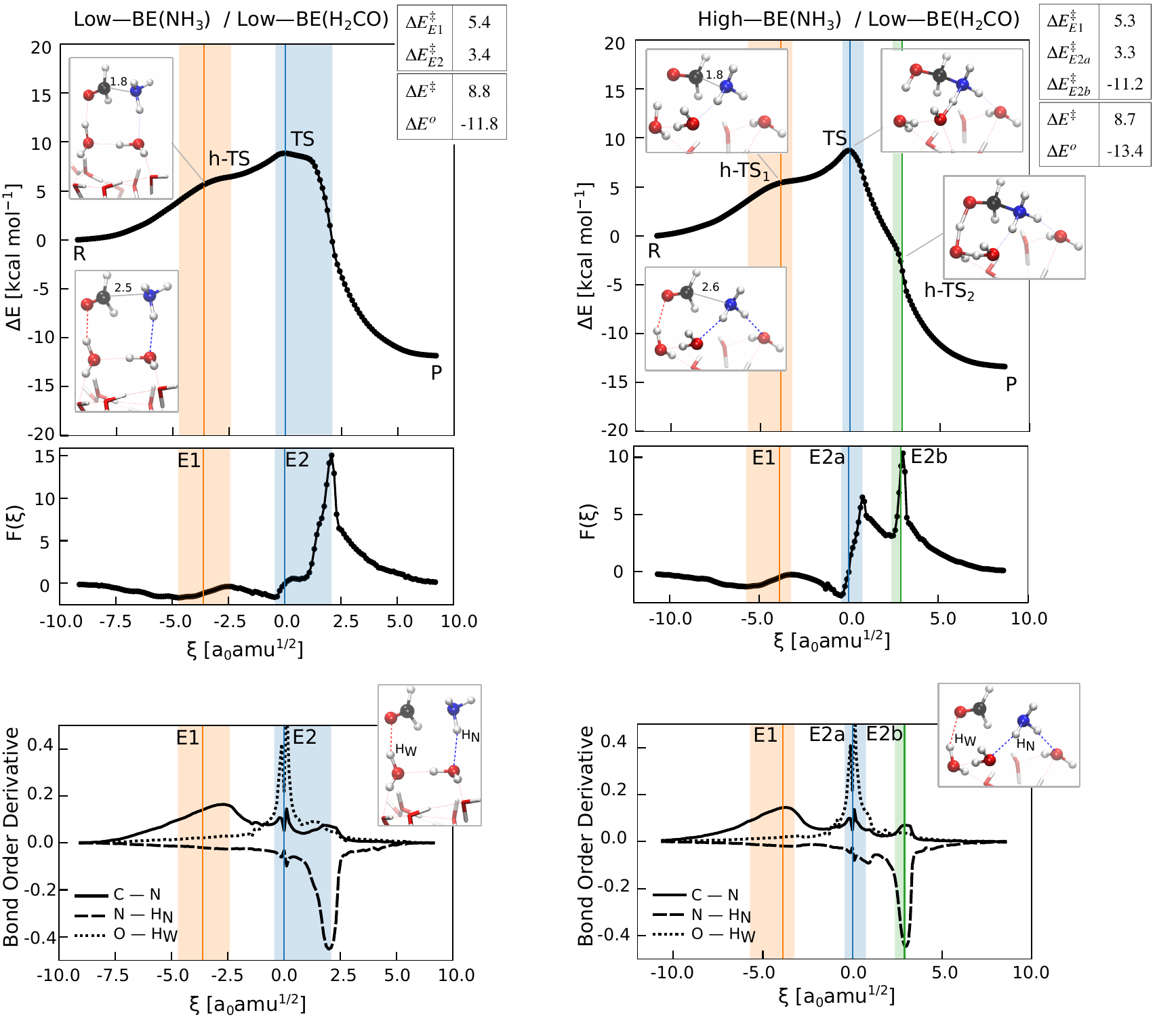}
     \caption{Left: energy (upper panel) and reaction force profiles (middle panel) along the intrinsic reaction coordinate ($\xi$) for system A, Low-BE(\ce{NH3)}/Low-BE(\ce{H2CO}) case. Energies are computed at BMK/def2-TZVP level of theory on geometries of BHandHLYP-D3BJ/def2-SVP quality. Blue and orange lines represent TS and h-TS, respectively. TS and h-TS regions are displayed as orange and blue shadowed areas. The inset table reports TS energy ($\Delta E^\ddagger$) and reaction energy ($\Delta E^o$) extracted from the energy profile, as well as the partition of the barrier in $\Delta E_{E1}^\ddagger$ (relative to h-TS) and $\Delta E_{E2}^\ddagger$ (TS). Inset figures representing R and h-TS are extracted from the IRC profile. C--N bond distances have been highlighted. The lower panel reports the bond order derivative for the main bond distances involved in Reaction \ref{eq:S1}. Right: analogous for system A, High-BE(\ce{NH3)}/Low-BE(\ce{H2CO}) case. The second hidden TS (\ce{h-TS_2}), present on the relaxation part of the reaction coordinate, is reported in green. The color scheme for the atoms is red for O, black for C, blue for N, and white for H. }
      \label{fig:irc_S1}
\end{figure*}

\noindent \textbf{\textit{Reaction mechanism:}} In order to elucidate the mechanism of Reaction \ref{eq:S1}, the IRC procedure is used to locate the reaction path.
Energy and reaction force profiles 
are reported in Fig.  \ref{fig:irc_S1}, for system A, which presents the lowest TS barrier.
The analysis of the reaction force profiles, middle panel, allows to define different reaction events that take place along the reaction coordinate.     
The TS region of the IRC profile is defined within the minimum and maximum of the reaction force 
profile and is displayed as a blue shadowed area. It has been pointed out\cite{toro-labbe_characterization_1999} that 
primarily electronic rearrangements occur in the TS region, whereas outside this region, structural 
modifications are predominant. However, additional inflection points in the reaction force profile 
suggest that a second incipient reaction event might be present 
before the TS region.
In fact,  the system presents an 
additional critical point, located before the TS,  that can be used to define a \textit{faux} TS. This event corresponds to a visible shoulder 
on the IRC profile. 
Such events  
 are considered \textit{transient} point along the reaction path and are associated with hidden transition states (marked as 'h-TS') and corresponding hidden intermediates ('h-I'), which can be converted into real TS and 
 intermediate in the presence of a change in the environment conditions or substitution pattern of the reactive fragments\cite{kraka_computational_2010}. Similarly to the TS region, a h-TS region can be defined as well, by means of the local minimum and maximum of the force profile (orange shaded region in Fig.  \ref{fig:irc_S1}).

In order to correlate the bond-breaking/forming processes of Reaction \ref{eq:S1} to the two events identified along the 
reaction path,
we analyzed the Wiberg bond orders\cite{weinhold_reduced_1967} and bond order derivatives along the reaction coordinate,  
for the main bonds involved in the reaction  (Fig.
\ref{fig:irc_S1}, lower left panel). 
 A negative sign in the derivative indicates bond weakening or dissociation, while a positive sign accounts for
bond formation or strengthening. 
We observed that the major change in the bond order describing the formation of C--N bond is located within the h-TS region, while the proton transfer processes are located in the TS region. 
Therefore, Reaction \ref{eq:S1} takes place in 
asynchronous fashion and presents 
two reactive events (E1 and E2): E1, \ce{C-N} bond formation, 
is the first to happen and gives rise to a dipolar h-I (\ce{^{-}OCH2}$\dotsb$\ce{NH3^{+}}),
which is then converted to the TS; 
followed by E2, the proton relay, which connects the TS to the product.
Information about the synchronicity of the chemical processes in E2 is also provided by the bond order derivative plot.  
Carbonyl-O protonation takes place first:  
\ce{O-H} forming peak is found to be at the beginning of the TS region, 
while \ce{N-H} breaking is closer to the end. 

\noindent \textbf{\textit{Partition of the reaction barrier:}}
Integration of the reaction force profiles allows the quantification of the energy associated with the different reactive events.   The total 
energy barrier for Reaction \ref{eq:S1}, $\Delta E^{\ddagger}$, is therefore partitioned as:
\begin{equation}
\Delta E ^{\ddagger} =  \Delta E_{{E1}}^{\ddagger} + \Delta E_{{E2}}^{\ddagger}
\label{eq:h_ts}
\end{equation}
where $\Delta E_{{E1}}^{\ddagger}$ is the energy barrier associated to formation 
of the h-TS and $\Delta E_{E2}^{\ddagger}$ is the energy necessary to convert 
the hidden intermediate to the TS.
The different values are shown in the inset of Fig. \ref{fig:irc_S1}, upper left.
The $\Delta E_{{E1}}^{\ddagger}$, associated to the h-TS,  
is 5.4 kcal
$\mathrm{mol^{-1}}$, while  $\Delta E_{{E2}}^{\ddagger}$, 
the barrier to 
convert h-I 
to the TS, is  3.4 kcal $\mathrm{mol^{-1}}$. 
The bond order analysis indicates that 
the barrier corresponding to $\Delta E_{{E2}}^{\ddagger}$  
is  mostly associated with the protonation of the carbonyl moiety 
of the dipolar adduct. 

In summary, structures that fall within the Low-BE(\ce{NH3})/Low-BE(\ce{H2CO}) binding regime exhibit a concerted mechanism,  featuring reaction barriers of approximately 10 kcal $\mathrm{mol^{-1}}$. However, analysis using 
IRC and reaction force indicates that the mechanism takes place in a highly asynchronous manner. Further breakdown of the reaction barrier reveals that the formation of the dipolar adduct represents the most significant energy expenditure in the TS barrier.
\subsubsection{High-BE(\ce{NH3)}/\/Low-BE(\ce{H2CO})}\label{sec:h_l}
This case exhibit structures in which \ce{NH3} is strongly bound:  
in addition to the water molecules involved in the proton relay,
there is another water molecule directly
coordinated to \ce{NH3}, serving as a secondary H-bond acceptor. Such additional H-bond interaction 
is congruent with the higher $IE(R)$, compared 
to the previous case sites, as displayed by the average value in Table \ref{tab:IE}.

We found  3 reactant configurations that correspond to this case,  
TS energy barrier and reaction energy results (systems A-C) are reported in   
Table \ref{tab:deltaE_S1}, and the energy diagram for the lowest energy path (system A) is in Fig. \ref{fig:S1_all_energy}, dashed grey line.  
The result, in terms of energy barrier, is analog to 
the previous case and to the model system, with a average barrier of 10 kcal $\mathrm{mol^{-1}}$. The reaction is exothermic by an average of -10 kcal $\mathrm{mol^{-1}}$.

\noindent \textbf{\textit{Reaction mechanism:}} A significant consequence of \ce{NH3} coordination
in the reactive complex manifests in the alteration of the reaction mechanism, evident when examining the IRC profile for structure A, displayed in   
Fig. \ref{fig:irc_S1}, right upper panel.     
While 
the left branch of the energy curve does not show major differences with respect to the previous case,
the curve in proximity of the region passed the TS presents the appearance of  another shoulder feature,  
corresponding to a second hidden transition state,  h-\ce{TS_2}(green solid line).
This is confirmed by the  analysis of the inflection points of the reaction force profile, Fig.   \ref{fig:irc_S1},
mid right panel, which displays  the presence of three 
distinct reactive events. The bond order derivative plot, Fig.  \ref{fig:irc_S1},
lower right panel, shows that the
asynchronicity in the proton transfer steps (E2) increases to such an extent that 
it manifests as two separated reactive events (E2a,b), in which the emerging h-TS$_2$ can be associated solely to the final step of the proton transfer: \ce{N-H} bond breaking.  
The result suggests it to be a peculiar feature of such reactive sites, following from the inductive effects exerted
by the two water molecules acting as H-bond acceptors on \ce{NH3}. They reduce the proton donor 
character of N-atom, thereby delaying the proton transfer from it. 
Basically, the effect exerted by the ice on \ce{NH3} side of the reactive complex enhances 
the asynchronicity of the proton transfer processes, hence \ce{N-H} breaking 
is located much later with respect to \ce{O-H} bond forming (E2a), and constitutes a separate event (E2b).\\
\noindent \textbf{\textit{Partition of the reaction barrier:}} The partition of the energy barrier  
shows that the reaction energy associated to E2a is 3.3 kcal $\mathrm{mol^{-1}}$,
which is similar to the Low-BE/Low-BE regime. As pointed out, the additional H-bond acceptor interaction affects the mechanism past the TS, therefore, it has a minor effect on the TS barrier. 

\subsubsection{Low-BE(\ce{NH3)}/\/High-BE(\ce{H2CO})}
This case includes reactive sites where  
\ce{H2CO} fragment is strongly bound. In terms of coordination, it corresponds to having an additional surface d(OH) 
 serving as a secondary H-bond donor on the carbonyl group. Such interaction exerts a significant effect on the reactive site, as reflected in the average $IE(R)$ for this group, amounting to 17.7 kcal $\mathrm{mol^{-1}}$ (Table \ref{tab:IE}).
We found four such reactive sites; 
Table \ref{tab:deltaE_S1} reports the corresponding TS energy barriers and reaction energies (systems A-D), and Fig. \ref{fig:S1_all_energy}, dashed black line, reports the energy diagram for system A.

\noindent \textbf{\textit{Reaction mechanism:}}  
Unlike previous cases, Reaction \ref{eq:S1} presents a step-wise mechanism, where 
the two steps, S(1,2), correspond to events E(1,2) previously identified: 
\begin{flalign}
      \text{S1} &&  \ce{H2CO} + \ce{NH3} \rightarrow \ce{^{-}OCH2}\dotsb\ce{NH3^{+}} &&
      \label{eq:S1_}
\end{flalign}
\begin{flalign}
      \text{S2} && \ce{^{-}OCH2}\dotsb\ce{NH3^{+}} \rightarrow \ce{HOCH2NH2}   &&
      \label{eq:S2_}
\end{flalign}
Once C--N bond is attained, the dipolar adduct is further stabilized by the two H-bond interactions established by the carbonyl-O  
and it emerges as a true stationary state.   

\noindent \textbf{\textit{Reaction barriers:}}
S1 is associated with the formation of a PES energy 
plateaux with 
an average energy barrier  
of 3.6 kcal $\mathrm{mol^{-1}}$ (before ZPVE correction, see Supplementary Material$\dag$, Table S3),
which is lower by about 2 kcal $\mathrm{mol^{-1}}$ than the partial barrier estimated for E1 in the Low-BE(\ce{H2CO}) regimes. The peculiar arrangement of \ce{H2CO} in the High-BE regime is associated with large intramolecular polarization of the fragment,
favoring the nucleophilic addition. 
S2, the conversion of the intermediate to the 
product, is exothermic and it is the rate-limiting step for all the systems, 
with a average \ce{TS_{S2}} energy of 7.5 kcal $\mathrm{mol^{-1}}$.
The number of water molecules involved in the proton relay (two for systems A, D and one for B,C) seems to 
have little effect on the TS energies, as both display cases 
with a higher and lower TS$_2$ energy. The variation  can rather be 
explained by analyzing the coordination of the reactive site to the 
ASW cluster. Taking into account not only the reactants but also the 
water molecules that participate in the proton relay, we found that
some systems (e.g. A and D) present
coordination defects in the H-bond ice network around the site, such as lacking H-bond donor groups acting on the 
assisting water molecules (see Supplementary Material$\dag$, Sec. S4). Energetically, 
it is an unfavorable characteristic since it prevents the stabilization of TS structures, increasing TS energies.    

\subsubsection{High-BE(\ce{NH3)}/\/High-BE(\ce{H2CO})}
The last group is constituted by structures which display the combination of the 
characteristics of the previous cases, reaching a number of four H-bond interactions present in the reactive complex (two H-bonds established by each 
reactant).
The level of insertion of the reactive complexes into the ice H-bond network is consequently higher, 
as also indicated by the $IE(R)$ values for this group, that are far greater in magnitude 
compared to the previous cases, with an average of 21.7 kcal $\mathrm{mol^{-1}}$ (Table \ref{tab:IE}). 
This group also presents the highest intramolecular polarization of both the reactants, as confirmed by the partial atomic charges computed on the main atoms involved in the reaction, see Supplementary Material$\dag$ Fig. S4, red, that show the largest values.
We found 2 structures with such characteristics. The results are illustrated in Table \ref{tab:deltaE_S1} (systems A-B) and Fig. \ref{fig:S1_all_energy} (system A).  

\noindent \textbf{\textit{Reaction mechanism:}} 
As for the previous case where \ce{H2CO} is a high-BE binding site, Reaction \ref{eq:S1} is comprised of two steps, S(1,2), which have been listed previously. 
Our findings suggest that it is indeed the \ce{H2CO} arrangement in the reactive site to determine the isolation of the dipolar intermediate, since such a feature is absent in the case where solely \ce{NH3} is in a high BE binding mode (Sec. \ref{sec:h_l}).

\noindent \textbf{\textit{Reaction barriers:}}
The main difference with previous case is that in the High-BE(\ce{NH3})/High-BE(\ce{H2CO}) regime,
\ce{TS_{S1}} energies are close to zero kcal $\mathrm{mol^{-1}}$, meaning that the PES is very flat and
the formation of the dipolar intermediate is essentially barrierless. 
In fact \ce{TS_{S1}} becomes negative for after ZPVE correction, whereas before the ZPVE correction the barrier is 0.1 kcal $\mathrm{mol^{-1}}$ (Table S3 in Supplementary Material\dag).
Such feature can be attributed to the fact that the reactive sites are especially suitable to undergo the reaction, as the ice surface contributes favorably to both the geometrical orientation of the fragments and in polarizing the bonds participating to Reaction \ref{eq:S1}, via inductive effects.  
Moreover, S1 involves a marked exothermic process (reaction energy $\Delta E_{S1}^{o}$ up to -6.2 kcal $\mathrm{mol^{-1}}$ ). This is imputed to the fact that, upon formation, the dipolar intermediate is strongly stabilized by the surface,
 as both sides of the adduct are coordinated to two water molecules.

Regarding S2, 
the proton transfer is the rate-limiting step of the process, as for the previous case, with barriers $\Delta E_{S2}^{\ddagger}$ ranging from 5 to 13 kcal $\mathrm{mol^{-1}}$. Despite the large barriers, it is possible
to argue that the exothermicity of S1 (especially for system A) might supply part of the amount
of energy needed to overcome S2.
\subsection{AMeOH formation inside of a nano-pore}\label{sec:S1_cavity}
To investigate the impact of the increased insertion of the reactive complex into the ice H-bond network, 
also motivated by the finding for Reaction  \ref{eq:S1} in proximity of  
high-BE sites, 
the reaction is conducted on a surface featuring a nano-pore.
\begin{figure*}
    \centering
    \includegraphics[width=0.9\textwidth]{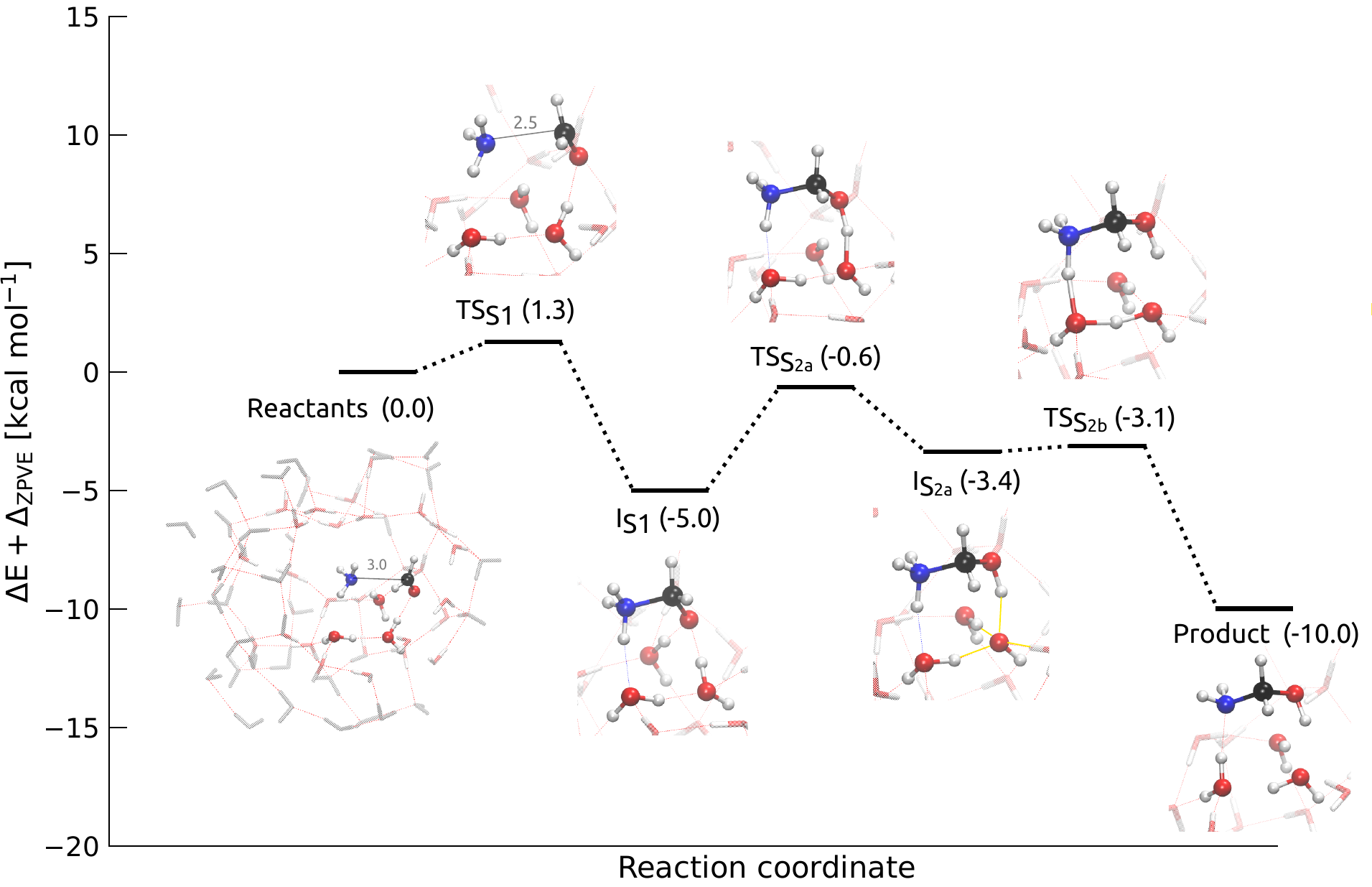}
     \caption{
     Energy diagram for Reaction \ref{eq:S1} inside a nano-pore. Energies values are computed at BMK/def2-TZVP // BHandHLYP-D3BJ/def2-SVP level of theory. Inset figures representing the energy minima are included. Water ice molecules directly coordinated to the reactants are represented as balls and sticks, while the rest as sticks. The molecules in the outer sphere of the pore are kept constrained during optimization and are represented as grey sticks. In the inset of \ce{I_{S2a}}, the donor H-bonds stabilizing the \ce{HO^-} in the dipolar adduct are highlighted in yellow. The color scheme for the atoms is red for O, black for C, blue for N, and white for H. }
      \label{fig:S1_porous}
\end{figure*} 
To attain the initial geometry of the reactive site, 
we sampled 
the inner region of the pore, 
applying a constraint 
to the cartesian positions of the  atoms belonging to the outer sphere, to preserve it. 
In the majority of the resulting reactant complexes, the docking site for the fragments was a specific d(OH) bond located in the central-right region of
the pore, thereby such configuration was the only one selected to carry out the reaction. An image of   
the reactive site equilibrium structure can be found in the inset of Fig. \ref{fig:S1_porous},  
which also reports the energy diagram and stationary point geometries for the process. 

\noindent\textbf{\textit{Reaction mechanism:}} 
We found a change in the reaction mechanism for Reaction \ref{eq:S1}, which appears to be  
comprised of three steps: S1, S2a and S2b.
While S1 corresponds to the very same step previously identified, i.e. the formation of the dipolar intermediate (\ce{I_{S1}}), 
the proton relay, S2, takes place in a different manner:    
\begin{flalign}
      \text{S2a} && \ce{^{-}OCH2}\dotsb\ce{NH3^{+}} + \ce{H2O} \rightarrow \ce{^{-}OH} \dotsb \ce{HOCH2NH3^{+}}  && 
\end{flalign}
\begin{flalign}
      \text{S2b} && \ce{^{-}OH} \dotsb \ce{HOCH2NH3^{+}}\rightarrow \ce{H2O} + \ce{HOCH2NH2}     && 
\end{flalign}
S2a, the first leg of the proton relay, leads to the isolation of a second dipolar intermediate (\ce{I_{S2a}}). This structure is analogous
to the second hidden dipolar intermediate (\ce{h-I_2}) that was detected as a shoulder on the energy profile in  
Fig.  \ref{fig:irc_S1}, upper right panel, green solid line,  for the High-BE(\ce{NH3}) case. 
The reason behind the isolation of the intermediate \ce{I_{S2a}}   
can be found in the
analysis of the ice H-bond network surrounding it.
Specifically, the water molecule that cedes the proton to the carbonyl group (first leg of the proton relay) 
presents 
a peculiar HB-coordination, being surrounded by four H-bond donor groups - highlighted in yellow in the inset of Fig. \ref{fig:S1_porous} -
which counterbalance the negative charge localized on the transient hydroxyl anion, contributing to the stabilization of \ce{I_{S2a}}. 
Therefore,  Reaction \ref{eq:S1} carried out inside of the nano-pore represents an example of a 'paused' mechanism, 
where the ice environment offers such H-bond coordination which allows to accommodate the reactive complex along the reaction coordinate, converting any of the events previously detected to actual reactive steps. 

\noindent \textbf{\textit{Reaction barriers:}}
The formation of C--N bond (S1) is exothermic and with an overall 
markedly low barrier (1.3 kcal $\mathrm{mol^{-1}}$), 
resembling the High-BE/High-BE case on ASW, reported in the previous section.   
The most notable difference across all studied systems on the 
set-of-clusters surface is that, in the case of the nano-pore, the 
transition state associated with the formation of the 
dipolar adduct (\ce{I_{S1}}) is practically  equal in energy with the 
proton  relay step. Furthermore,  the latter is split into two steps 
with an additional intermediate, in which the TS barriers of the 
protonation of the carboyl group (\ce{TS_{S2a}}) is higher
than that of the dissocation of the N-H bond to form the amino moiety 
(\ce{TS_{S2b}}) 
In summary, Reaction \ref{eq:S1} carried out in a realistic ASW pore presents 
a low energy pathway in which the highest energy on the reaction path amounts
to only 1.3 kcal $\mathrm{mol^{-1}}$ and the intermediates are all 
exothermic. Therefore, such a 3-step variation of the reaction mechanism for 
Reaction \ref{eq:S1} seems to be the most plausible to take place under 
interstellar conditions.

\begin{table*}[h]
    \centering   
\caption{ 
Reaction \ref{eq:S1} carried out on different ASW sites and inside of a nano-pore,  using BHandHLYP-D4/def2-SVP geometries, 
     computed at $\omega$-B97M-D3BJ/def2-TZVP level of theory, for reaction on ASW, and 
     BMK/def2-TZVP, for the nano-pore. Column two indicates the binding regime of the reactants; column three reports the systems for each case, ordered alphabetically from the lowest TS energy, column four indicates the number X of water molecules involved in the proton relay. The rest of the columns report energy barriers ($\Delta{E_{En}}^{\ddagger}$) and reaction energies ($\Delta{E_{En}}^o$), calculated with respect to the energy of the reactants and including ZPVE correction,  for step \textit{n} of the reaction, if present, along with the overall TS and reaction energy ($\Delta{E}^{\ddagger}$  and $\Delta{E}^o$, respectively).  
          Values in kcal $\mathrm{mol^{-1}}$.} 
\begin{tabular*}{\textwidth}{@{\extracolsep{\fill}}llllllllllll}
    \hline
                AMeOH formation & & \# & \ce{W_X} & $\Delta E_{S1}^{\ddagger}$ & $\Delta E_{S1}^{o}$ & 
          $\Delta E_{{S2a}}^{\ddagger}$ & $\Delta E_{{S2a}}^{o}$ & $\Delta E_{{S2b}}^{\ddagger}$ & $\Delta E_{{S2b}}^{o}$ & $\Delta E^{\ddagger}$ & $\Delta E^{o}$\\
\hline
\textit{ASW clusters:}         &&&&&&&&&\\
&  Low-BE(\ce{NH3)}/\/Low-BE(\ce{H2CO})   
     &    A  & 2 & &  & &&&&  9.6  & -7.4  \\
&      &   B  & 2 &  &  & &&&&10.1 & -7.5  \\
&      &   C  & 2 &  &  & &&&&10.2 & -7.3 \\
&      &   D  & 2 &  &  & &&&&10.3 & -7.8 \\
&      &   E  & 1 &  &  & &&&&14.3& -12.9  \\
                               
&High-BE(\ce{NH3)}/\/Low-BE(\ce{H2CO}) 
&        A  & 2 & &&&& && 9.6 & -5.8  \\
&&        B  & 2 & &&&& && 10.1 &  -8.1 \\
&&        C  & 2 & &&&& && 10.2 & -6.7 \\
&        Low-BE(\ce{NH3)}/\/High-BE(\ce{H2CO})
 &        A  & 2 & 5.2 & 5.1 & 7.5 & -9.1  &&& 7.5 & -9.1\\
&&         B  &  1 & 5.2 & 5.1 & 11.6 & -8.3  &&& 11.6 & -8.3 \\
&&         C  & 1 & 5.9 & 6.2 & 9.1 & -8.1 &&& 9.1 &  -8.1 \\
&&         D & 2   & 8.1 & 8.3 &12.5 & -7.6 & && 12.5 & -7.6\\
&High-BE(\ce{NH3)}/\/High-BE(\ce{H2CO}) &              
 A  & 2  & -0.6 & -6.2 & 1.5 & -11.5 &&& 1.5 &  -11.5\\
&& B  & 1 &  1.5 & -0.1 &  5.1 &  -9.9 & & & 5.1 &  -9.9\\\\
        \textit{Porous ASW:} & &A & 2 & 1.3 & -5.0 &  -0.6 & -3.4 & -3.1 &  -10.0 & 1.3 & -10.0\\    
    \end{tabular*}
    \label{tab:deltaE_S1}
\end{table*}

\begin{figure}[t]
    \centering
    \includegraphics[width=0.48\textwidth]{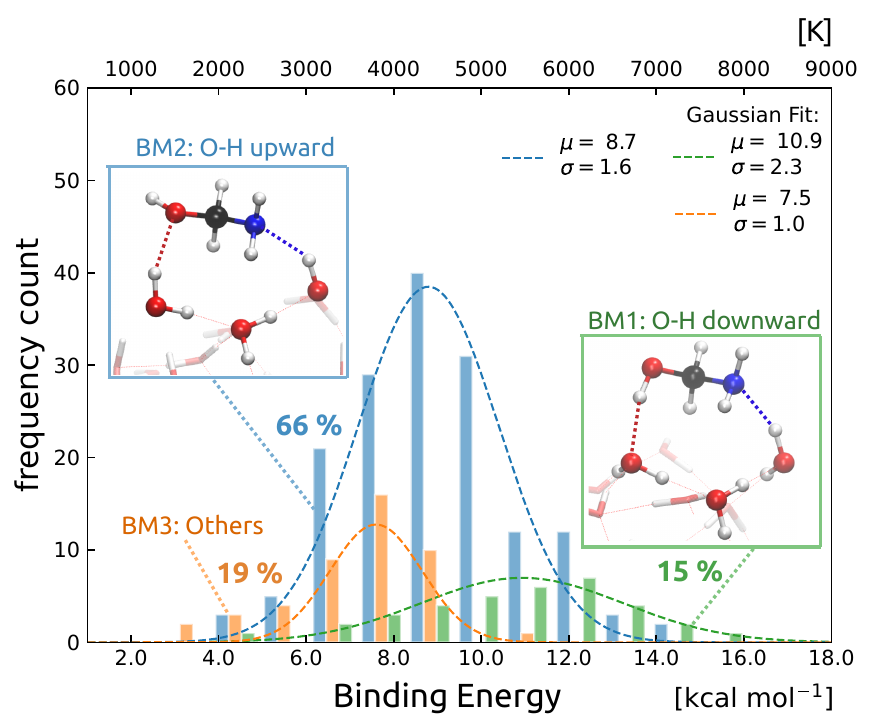}
	\caption{Histogram of the BE distribution of AMeOH, calculated at $\omega$PBE-D3BJ/def2-TZVP//BHandHLYP/def2-SVP  level of theory, without including ZPVE correction. 
 Mean ($\mu$) and
standard deviation ($\sigma$) of a Gaussian fit are reported for the main binding modes identified (BM1, where the hydroxyl moiety of the molecule acts as H-bond donor, and BM2, where it acts as H-bond acceptor), along with an example, as selected from the distribution. The color scheme
for the atoms is red for O, grey for C, white for H and blue for N.}
     \label{fig:be_modes_aminomethanol}
\end{figure} 

\subsection{Binding Energy distribution of AMeOH}
AMeOH can  
establish a variety of interactions with the ASW, owing to the amino and hydroxyl groups 
that can both act as H-bond donors and acceptors. 
Fig. \ref{fig:be_modes_aminomethanol} displays the BE 
distribution of AMeOH, which is spread over 10 kcal $\mathrm{mol^{-1}}$. 
A large molecule, such as AMeOH,  presents several binding modes;  a complete census of them is out
of the scopes of this work.
We identified two majority modes;  
the main difference between them is that in the first binding mode (BM1, green), 
the hydroxyl OH-group acts as H-bond donor, whereas in the second (BM2, blue), the hydroxyl 
accepts a H-bond from a d(OH) surface group. BM1 exhibits a higher average 
BE (10.9 kcal $\mathrm{mol^{-1}}$)  than BM2, despite being rarer (only 15 \% of the binding sites). 
Interestingly,
AMeOH formation pathways explored in this 
work result in AMeOH binding modes corresponding to the high-energy structures in BM1 (as they derive from the protonation of carbonyl-O). This indicates that, once formed through Reaction \ref{eq:S1}, AMeOH is expected to be bound very strongly to the ASW.

\section{Discussion}\label{sec:discussion}

The detailed analysis of the nucleophilic addition of \ce{NH3} to \ce{H2CO}, 
resulting in AMeOH, revealed that the presence of an ASW surface leads to a wide 
variety of reaction mechanisms.  These span from a single concerted reaction 
exhibiting various degrees of asynchronicity, to a total separation of
the three principal reaction events that result in the formation of AMeOH:
the C--N bond formation to establish the AMeOH backbone, 
the protonation of the carbonyl group (O--H bond formation), 
and the N--H bond dissociation that gives rise to the amino group.
Furthermore, such diversity significantly affects the overall transition 
state energies,  reducing them from $\sim$10 kcal $\mathrm{mol^{-1}}$ 
to as low as 1.3 kcal $\mathrm{mol^{-1}}$. 

The C--N bond formation involves a dipolar adduct (\ce{^{-}OCH2}$\dotsb$\ce{NH3^{+}}), 
which is present in all the reaction pathways. In the concerted reactions
it is a hidden intermediate, but becomes a discernible stationary state
as the interaction energy between the ice surface and the reactive complex increases. 
In the High-BE(\ce{NH3})/High-BE(\ce{H2CO}) regime and inside of the nano-pore, 
the large number of
water molecules favorably  coordinated to the reactants, lowers the energy of the reactive complex, as well as of the TS and dipolar intermediate, 
to such extent that this step
 (S1) 
is rendered practically barrierless (between -0.6 and 1.5 kcal $\mathrm{mol^{-1}}$).
Moreover, in those limiting cases,  
the formation of the intermediates is notably  exothermic (range of 3--6 kcal $\mathrm{mol^{-1}}$). 
Evidence of such dipolar intermediate can be found in a DFT study using PCM \cite{chen_theoretical_2011}, 
where the reaction proceeds on a 4-water-molecules cluster. 
In the study, however, the conditions under which the intermediate 
was isolated, were not investigated, as it formed barrierless. Comparing 
with the results reported herein, it is likely that the addition of implicit solvation effects  overestimates the actual ASW stabilization on the 
dipolar adduct. 

The second part of the reaction is constituted by proton transfer processes. 
In the concerted pathways they are present as distinct events along the reaction profile, 
while in the High-BE(\ce{H2CO}) cases, they represent the second step of a step-wise mechanism. 
Nevertheless, only in one case (A) of the High-BE(\ce{NH3})/High-BE(\ce{H2CO}) regime, 
the \ce{TS_{S2}} energy is significantly lowered (1.5 kcal $\mathrm{mol^{-1}}$) due to the hyper-coordination through H-bonds on the water molecules that channel the proton relay. Hence, it is worth noting that
the first TS  (\ce{TS_{S1}}) is stabilized by the H-bond network provided  by water molecules coordinated to the reactive complex, while the second TS (\ce{TS_{S2}}) is stabilized by the H-bonds established by the water molecules assisting in the proton relay, i.e. the molecules present in the lower ice layers.  
When the reaction takes place inside a nano-pore,  this effect is amplified, such that the proton relay 
is further divided into two distinct stationary states, relating to first the protonation 
of the carbonyl moiety, followed by  N--H bond dissociation in \ce{NH3}, 
with the new intermediate (\ce{I_{S2a}}) showing exothermic character. 
Another aspect that influences the height of the energy barriers 
is the number of molecules involved in the proton relay step of the
reaction. 
\citet{baiano_gliding_2022} studied the \ce{HNC <=> HCN} reaction 
with different sized cluster models and found  that
the isomerization barrier was lower when four water molecules where involved 
in the proton-relay. As we only searched for reaction paths involving 1 and 2-water 
molecules, it is possible for a mechanism involving more water molecules
to further lower the reported TS barriers for the Low-BE/Low-BE cases.
However, we expect less dependence on the number of water molecules for this reaction,  since 
the IRC analysis showed that 
the major part of the reaction barrier 
corresponds to the addition of  the two fragments and that only the first 
leg of the proton relay contributes  to the barrier.

In light of these results, it seems apparent that, overall, 
the prerequisite for a 
low energy mechanism  
is linked to having a high interaction energy between the reactive complex and the ice, driven by the specific arrangement of the \ce{H2CO} moiety,   
which requires the presence of a pair of d(OH) bonds. 
In order to estimate the frequency of occurrence of these pair-d(OH)s bonds, 
we quantified these particular sites on the large periodic ASW surface models (see Fig. \ref{fig:asw_models}). We took into account 
five periodic ASW surfaces that stem from the same MD simulation, to get a statistically 
relevant quantification.  The census  considered only the surface of the periodic 
slabs i.e. the molecules at the top layer. 
Assuming an adsorption site to have an area of approximately $3 \times 3 $ \text{\AA},
each periodic surface is estimated to accommodate $\sim$100 potentially unique reactive sites, for a total of 500 possible reactive sites. 
The result of this analysis indicates that 25\% over the 500 sites are estimated to 
harbor d(OH) bonds, on average over the five surfaces. It is worth mentioning that the number  
includes d(OH)s that are located in concavities or pores, as those regions constitute 
part of the periodic slabs. Finally, pair-d(OH)s sites - two d(OH) sites in close 
proximity, below 3 \text{\AA} from one another - are present in approximately 5\%  of all possible reactive sites: a 
non-negligible number, thus making the low energy mechanism feasible over the vastness 
of a real ice-grain surface.

\subsubsection*{Astrophysical implications}
One of the main results of this work is that,  
under specific  circumstances,
concerted water-assisted reactions are converted to step-wise.
Specifically, 
a dipolar-adduct is isolated in the first part of Reaction \ref{eq:S1}, 
and the proton relay, taking place in the second part, is fragmented to a sequence of isolated events.    
As a consequence, the rate of the  proton transfer processes might benefit from 
quantum tunneling effects, allowing reactions involving hydrogen atoms to
occur faster than expected from transition state theory. 
Tunneling effects associated with transfer and abstraction reactions involving hydrogen, 
have been extensively studied\cite{hasegawa_new_1993,tielens_physics_2005,goumans_isotope_2011} 
and are known to play a significant role in the ISM\cite{song_formation_2016,song_tunneling_2017}. Moreover, tunneling effects are heavily 
dependent on the width of the energy barrier. Analysis of the IRC profiles revealed that, 
in case of isolated or nearly isolated proton transfer steps, e.g. High-BE(\ce{NH3})/Low-BE(\ce{H2CO}) regime, the top region of the IRC curve, that corresponds to the proton 
transfer steps, gets particularly narrow, compared to the profile of the Low-BE/Low-BE case (where the process involves the entire set of proton relay steps)   
opening to the actual    
possibility 
of tunneling effects. 

An additional effect, that could influence the outcome of the reaction
under investigation in astronomical environments, is the reaction energy
dissipation rate to the ice matrix. In some limit cases that present 
exothermic intermediates, the viability of the reaction will greatly depend 
on the competition between thermalization and continuation of the reaction.
Estimating the timescales for this competition is 
challenging, 
and dedicated MD
simulations can shed light in this particular topic. Nevertheless,
Fig. \ref{fig:S1_porous} shows that, inside a nano-pore, the reaction should definitely take place, due to the overall small barriers, which summed to the efficiency of quantum tunneling in the proton relay mechanism (\textit{vide supra}), renders thermalization of I$_{\text{S1}}$ or I$_{\text{S2a}}$ on the surface an unlikely event.

Our results support the proposition that AMeOH  can be formed through the initial reaction of a  
Strecker-type synthesis under interstellar dense clouds condition.  Furthermore, the results 
presented in this  work are in excellent agreement with the 
experimental result \citet{bossa_nh_2009} of 
0.5 kcal $\mathrm{mol^{-1}}$ for the energy barrier;  especially considering  the 
rich diversity of catalytic sites present on ASW, which suggests that TS 
energies smaller than our lower limit (1.3 kcal $\mathrm{mol^{-1}}$) might be encountered. Furthermore,  
our work lends support to the claim that AMeOH is present 
on the ice, and has a long residence time, in light of the range of 
BEs we computed (whose upper limit of 15 kcal $\mathrm{mol^{-1}}$, is in good 
agreement with the experimental result by \citet{bossa_nh_2009}  of 14 kcal $\mathrm{mol^{-1}}$). 
Possible depletion routes of AMeOH include the second step of the Strecker synthesis, 
namely the dehydration to form methynimine. However, this reaction has been found to have
a significant energy  barrier\cite{rimola_can_2018}.  Moreover, the orientation of AMeOH of ASW as synthesized \textit{via} Reaction \ref{eq:S1} does not enable the dehydration, 
making it unlikely that the exothermicity of Reaction \ref{eq:S1} 
could be carried over, to overcome the 
dehydration barrier under cold interstellar cloud conditions. Based on the presented results, it seems more 
likely that the non-detection of AMeOH  
is related to the identification difficulties. 
Infrared astronomical spectra of AMeOH display strong absorption features, which are blended with
water and silicate bands. Furthermore the remaining bands fall in regions where several 
other organic compounds absorb.   
Future  surveys 
with JWST might help bridge the gap between experiments and observations, although more detailed 
studies on both the IR spectrum of AMeOH and its mm rotational features in the gas phase are 
likely needed. These studies could facilitate the detection of this prebiotic precursor using ALMA, in regions 
where warming from newborn stars can release AMeOH into the gas phase.

\section{Conclusions}

In this paper, we presented a broad analysis of the catalytic effect
of an ASW surface for the reaction of \ce{NH3} and \ce{H2CO}
to form \ce{NH2CH2OH}.  We studied the reaction on a set of different 22-water-molecules clusters 
and inside a nano-pore, derived from a  periodic ASW model surface  containing 
500 water molecules, generated according to the initial density of 0.8 g $\mathrm{cm^{-3}}$. We sampled four catalytic sites with an
approach  based on the binding modes derived from the binding energy 
distribution of \ce{NH3} and \ce{H2CO}  molecules, and determined that the catalytic sites can be classified into four groups based on the interaction 
of the carbonyl-end the amino group with the ice surface, in the reactant
complex, namely 
Low-BE(\ce{NH3)}/Low-BE(\ce{H2CO}), 
High-BE(\ce{NH3)}/Low-BE(\ce{H2CO}), 
Low-BE(\ce{NH3)}/High-BE(\ce{H2CO}), 
High-BE(\ce{NH3)}/High-BE(\ce{H2CO}). The findings about the catalytic effect of 
the ASW surface can be summarized as follows:

\begin{enumerate}
\item In the cases where the interaction between the reaction complex and  the 
ASW surface is weak - Low-BE(\ce{NH3)}/Low-BE(\ce{H2CO}) regime -  the TS energy is in the range of 9.6-14.3 kcal $\mathrm{mol^{-1}}$,
the reaction takes place in a one-step concerted mechanism. The 
energy barriers are similar to the reaction carried out on a water dimer model surface, 
indicating a negligible role of the bulk water on the reaction mechanism. 
However, the analysis of IRC, reaction force, and NBO bond orders, allowed
us to classify the mechanism of
the reaction
as highly asynchronous: the C--N bond 
formation is present as a hidden-TS in the form of an emerging dipolar adduct, 
which is 95\% completed before the proton relay begins. 

\item When the interaction with the ASW surface increases through the \ce{NH3}-end of the reactant complex - High-BE(\ce{NH3)}/Low-BE(\ce{H2CO}) - the reactions is still 
concerted, however, the reaction mechanism becomes more asynchronous as
both proton transfer events that constitute the proton relay are now 
present as distinct hidden-TSs.

\item When the interaction with the ASW surface increases through the carbonyl-end of the reactant complex - Low-BE(\ce{NH3)}/High-BE(\ce{H2CO}) - the reactions transits 
from being concerted with a hidden-TS to displaying a well-characterized 
dipolar intermediate for the C--N bond formation, albeit within a flat 
potential energy region. Furthermore, the  highest TS energy, 
that corresponds to the TS of the proton relay, is only slightly lower than 
the single TS energy of the concerted mechanism of the previous cases.

\item In the case in which both ends of the reactant complex are bound 
to the ASW surface in a high BE configuration - High-BE(\ce{NH3)}/High-BE(\ce{H2CO}) -  the effect on the TS 
barriers is most significant, resulting in the nearly barrierless 
formation of a 
exothermic dipolar intermediate.  
In the most effective catalytic site of this interaction type, the highest TS 
energy  is also considerably lower than in all the previous
cases (1.5 kcal $\mathrm{mol^{-1}}$) and corresponds to the proton relay step of the reaction.

\item  When the reaction takes place in a nano-pore catalytic site, the highest TS energy corresponds
to the first leg of the proton relay and display a low value (1.3 kcal $\mathrm{mol^{-1}}$), within the range 
of the High-BE/High-BE case. However, the mechanism presents an 
additional fragmentation, 
as the proton relay is separated into 
two exothermic steps for the N--H bond dissociation and the O--H bond formation.
The emergence of these intermediates can be attributed 
to the nano-pore environment, which provides the stabilization of the anionic hydroxyl 
intermediate in the proton relay chain, thanks to the water molecules in the neighboring ice layers.

\item The d(OH) and pair of d(OH) bonds on the ASW surface are essential 
for the reaction to occur. A survey of these bonds across the possible binding sites on 
five different 500-water-molecules ASW periodic surfaces, revealed that d(OH) and pair d(OH) bonds are 
present at a fraction of 25\% and  5\%, respectively.

\item Analysis of the main adsorption motives in the BE distribution of AMeOH on the set-of-clusters, shows that the orientation of the product, as generated through Reaction \ref{eq:S1}, corresponds to the binding mode of highest BE. 
\end{enumerate}

\noindent Our findings corroborate the hypothesis that AMeOH can be 
synthesized through the initial reaction of a Strecker-type synthesis 
under interstellar dense cloud conditions, aligning with experimental
results that estimated a 
low activation barrier. 
Furthermore, the presence of diverse catalytic 
sites on ASW  suggests even lower transition state energies are feasible.  
Finally, the binding energy distribution of AMeOH indicates a long residence time on ice, 
supporting the idea that the non-detection of AMeOH in the ISM may stem more from 
identification difficulties than from its absence. To facilitate detection, 
new experimental and theoretical spectroscopic data are needed both on the ice 
and in the gas phase, which could open avenues for detecting this molecule with both JWST and ALMA. The effect of the ASW surface on the  subsequent steps of the Strecker Synthesis of Glycine, following from AMeOH formation, will be addressed in a forthcoming work.

\section*{Conflicts of interest}
There are no conflicts to declare.

\section*{Acknowledgements}
GMB gratefully acknowledges support from ANID Beca de Doctorado Nacional 21200180 and Proyecto UCO 1866-Beneficios Movilidad 2021-2022.
GSV thanks the Universidad de Concepción for the Beca de Excelencia Académica scholarship. SB acknowledges BASAL Centro de Astrofisica y Tecnologias Afines (CATA), project number FB210003.
GM and JK thank the Deutsche Forschungsgemeinschaft (DFG, German Research Foundation) for supporting this work by funding - EXC2075 – 390740016 under Germany's Excellence Strategy and acknowledge the support by the Stuttgart Center for Simulation Science (SimTech). GM acknowledges the support of the grant RYC2022-035442-I funded by MCIN/AEI/10.13039/501100011033 and ESF+. 
SVG thanks VRID research grant 2022000507INV for financing this project.

\balance

\bibliography{Astrochem,Libros_FQ,PhD_Project,QCMM,extra_ref} 
\bibliographystyle{rsc}

\end{document}